\documentclass[twocolumn,english,showpacs,preprintnumbers,amsmath,amssymb,floatfix]{revtex4-1}

\usepackage[T1]{fontenc}
\usepackage[latin9]{inputenc}
\usepackage{color}
\usepackage{array}
\usepackage{amstext}
\usepackage{graphicx}
\usepackage{esint}
\usepackage{rotating}
\usepackage{appendix}

\usepackage[font=small,labelfont=bf]{caption}
\usepackage{xcolor}
\usepackage[colorlinks = true,
linkcolor = magenta,
urlcolor  = blue,
citecolor = red,
anchorcolor = blue]{hyperref}

\makeatletter

\providecommand{\tabularnewline}{\\}

\@ifundefined{textcolor}{}
{%
 \definecolor{BLACK}{gray}{0}
 \definecolor{WHITE}{gray}{1}
 \definecolor{RED}{rgb}{1,0,0}
 \definecolor{GREEN}{rgb}{0,1,0}
 \definecolor{BLUE}{rgb}{0,0,1}
 \definecolor{CYAN}{cmyk}{1,0,0,0}
 \definecolor{MAGENTA}{cmyk}{0,1,0,0}
 \definecolor{YELLOW}{cmyk}{0,0,1,0}
 }

\@ifundefined{definecolor}
 {\usepackage{color}}{}
\@ifundefined{definecolor}
 {\usepackage{color}}{}
\makeatother
\usepackage{babel}

\begin{document}




\title {Nucleon spin structure functions at NNLO in the presence of target mass corrections and higher twist effects}

\author { Hamzeh Khanpour$^{1,2}$ }
\email{Hamzeh.Khanpour@mail.ipm.ir}

\author {S. Taheri Monfared$^{2,3}$ }
\email{Sara.Taheri@ipm.ir}

\author {S. Atashbar Tehrani$^{4}$ }
\email{Atashbar@ipm.ir}

\affiliation {
$^{(1)}$Department of Physics, University of Science and Technology of Mazandaran, P.O.Box 48518-78195, Behshahr, Iran        \\
$^{(2)}$School of Particles and Accelerators, Institute for Research in Fundamental Sciences (IPM), P.O.Box 19395-5531, Tehran, Iran      \\
$^{(3)}$Department of Physics, Faculty of Basic Science, Islamic Azad University Central Tehran Branch (IAUCTB), P.O. Box 14676-86831, Tehran, Iran \\
$^{(4)}$Independent researcher, P.O. Box 1149-8834413, Tehran, Iran   }

\date{\today}

%
%
\begin{abstract}\label{abstract}

We extract polarized parton distribution functions (PPDFs), referred to as ``KTA17,'' together with the highly correlated strong coupling $\alpha_s$ from recent and up-to-date $g_1$ and $g_2$ polarized structure functions world data at next-to-next-to-leading order in perturbative QCD.
The stability and reliability of the results are ensured by including nonperturbative target mass corrections as well as higher-twist terms  which are particularly important at the large-$x$ region at low Q$^2$. Their role in extracting the PPDFs in the nucleon is studied. Sum rules are discussed and compared with other results from the literature. This analysis is made by means of the Jacobi polynomials expansion technique to the DGLAP evolution.
The uncertainties on the observables and on the PPDFs throughout this paper are computed using standard Hessian error propagation which served to provide a more realistic estimate of the PPDFs uncertainties.

\end{abstract}

\pacs{13.60.Hb, 12.39.-x, 14.65.Bt}
\maketitle
\tableofcontents{}

%
%
\section{Introduction}\label{Introduction}

Hadrons are the complex systems consisting of quarks and gluons. The determination of parton densities and understanding the details of their $x$ and $Q^2$ dependence is one of the most important challenges in high energy physics.
A straightforward calculation of the cross section is available via the collinear factorization theorem in perturbative QCD (pQCD).	
Particularly interesting is the investigation of polarized processes which provides information about the basic decomposition of nucleon's spin into its quark and gluon constituent parts.
In recent years, the deep inelastic scattering (DIS) of polarized leptons off polarized nucleons has played an important role
in the study of the nucleon spin structure functions.	
The spin structure of the nucleon is still one of the major unresolved issues in the study related to hadronic physics~\cite{Aidala:2012mv}. While the combined quark and antiquark spin contributions to the nucleon spin, have been measured to be about 30\%, the contribution of the gluon spin to the spin of the nucleon is still insufficiently constrained after more than two decades of intense study.
The last few years have witnessed tremendous experimental and phenomenological progress in our understanding on the spin structure of the nucleon.
There are several QCD analyses of the polarized DIS
data along with the estimation of their uncertainties in the literature \cite{Ball:2016spl,deFlorian:2009vb,Hirai:2008aj,Blumlein:2010rn,Leader:2010rb,Nocera:2014gqa,Jimenez-Delgado:2013boa,Jimenez-Delgado:2014xza,Sato:2016tuz,Arbabifar:2013tma,Monfared:2014nta,Shahri:2016uzl,Gluck:2000dy}.

Current phenomenological spin-dependent parton distribution function (PDF) analysis uses the spin-dependent DIS measurements on $g_1^{\rm p, n, d}(x, Q^2)$ and $g_2^{\rm p, n, d}(x, Q^2)$, see Table~\ref{tab:DISdata}. Beside these data sets, one can also include the recent PHENIX measurement on neutral-pion $\pi^0$ productions~\cite{Adare:2014hsq,Adare:2015ozj} at $\sqrt{s}=200, \, 510$ GeV and inclusive jet production from the STAR Collaboration~\cite{Adamczyk:2014ozi} in polarized proton-proton collisions at the Relativistic Heavy Ion Collider (RHIC). The longitudinal single-spin asymmetries in $W^\pm$ weak boson production~\cite{Adamczyk:2014xyw,Adare:2015gsd} from polarized proton-proton collisions also can be used. These data sets may lead to better determination of the polarized gluon, sea quark and antiquarks distributions at small-$x$.

The precision of polarized parton distribution functions (PPDFs) determination in QCD analyses has steadily improved over the recent years, mainly due to refined theory predictions for the hard parton scattering reactions and also more accurate experimental observables.
Recently, the COMPASS Collaboration at CERN~\cite{Adolph:2015saz} extracted the spin-dependent structure function of the proton $g_1^p(x, Q^2)$ and the longitudinal double-spin asymmetries $A_1^p (x, Q^2)$ from scattering of polarized muons off polarized protons for the region of low $x$ (down to $0.0025$) and high photon virtuality $Q^2$.
Although significant progress has been made, the gluon polarization, as a fundamental ingredient describing the inner structure of the nucleon, suffers from large uncertainties and remains poorly constrained. Worse even, the gluon distributions originating from different collaborations represent significant differences.

In our latest analysis TKAA16~\cite{Shahri:2016uzl}, we performed the first detailed pQCD analysis using Jacobi polynomial approach at next-to-leading-order (NLO) and next-to-next-to-leading-order (NNLO) approximation.
All the available and up-to-date $g_1^{\rm p, n, d}(x, Q^2)$ world data including recent COMPASS measurements~\cite{Adolph:2015saz} were considered which led to new parametrization of spin-dependent parton destinies.
In the discussed paper~\cite{Shahri:2016uzl}, we simply considered the equality of $g_1(x, Q^2) \equiv g_1^{\tau2}(x, Q^2)$, while the information on quark-gluon correlation is encoded into the higher-twist parts of the $g_1(x, Q^2)$ and $g_2(x, Q^2)$. Here, $\tau2$ means twist 2 and HT refers to higher twist. Although these dynamical effects are suppressed by inverse powers of $Q^2$ in the HT expansion of $g_1(x, Q^2)$, they appear to be equally important as their twist-2 part in $g_2(x, Q^2)$.
This special property makes measurements of $g_2(x, Q^2)$ particularly sensitive for investigating multiparton correlations in the nucleon.
Furthermore, $g_2(x, Q^2)$ observables are mostly in the low $Q^2$ region where the target mass corrections (TMCs) and HT effects become significant. In the current analysis, which we refer to as ``KTA17,'' we develop a precise analysis by including TMCs and HT contributions in both $g_1(x, Q^2)$ and $g_2(x, Q^2)$ structure functions.
The role of these corrections in PPDFs estimation using pQCD fits to the data is discussed.
Studies of the moments of spin-dependent structure functions provide an opportunity to test our understanding of pQCD like that of the Bjorken sum rule.
We also demonstrate once more the reliability and validity of the Jacobi polynomial expansion approach at the NNLO approximation to extract the PPDFs from polarized DIS structure function.

The remainder of this article is organized as follows:
In Sec.~\ref{Theoretical-framework}, we review the theoretical formalism underpinning the KTA17 analysis of the polarized DIS structure function,  the Jacobi polynomials approach, target mass corrections and higher-twist effects.
Section~\ref{global-PPDFs} provides an overview of the method of the analysis, data selection, $\chi^2$ minimization and error calculation.
The results of present NNLO polarized PDFs fits and detailed comparison with available observables are discussed in Sec.~\ref{Results-of-PPDFs}.
We compute and compare associated polarized sum rules in
Sec.~\ref{Sum-Rules}. A short discussion on the present status of polarized PDFs global analyses is discussed in Sec.~\ref{high-precision-era}. Finally, Sec.~\ref{Summary} contains the summary and concluding remarks.
In Appendix~\ref{AppendixA}, we present a FORTRAN package containing results for the KTA17 polarized structure functions at NNLO approximation together with corresponding uncertainties. Appendix~\ref{AppendixB} provides the analytical expressions for the polarized NNLO quark-quark and gluon-quark splitting functions.

%
%
\section{Theoretical framework}\label{Theoretical-framework}

In this section, we review the basic theoretical framework for the polarized DIS structure functions on which the  KTA17 PPDFs analysis is based.
After a brief revision of the leading-twist structure functions at NNLO approximation, we present the Jacobi polynomials expansion method which was already used to extract KTA17 PPDFs at NNLO approximation from polarized DIS data ~\cite{Shahri:2016uzl}. Our approach to take into account TMCs and HT corrections is discussed in the following subsections.

\subsection{Leading-twist polarized DIS structure function}\label{structure-function}
%

In the light-cone operator-product expansion (OPE), the leading-twist (twist $\tau =2$) contributions correspond to scattering off asymptotically free partons, while the higher-twist contributions emerge due to multiparton correlations.
The leading-twist spin-dependent proton and neutron structure functions, $g_1^{\rm p,n} (x, Q^2)$ at NNLO, can be expressed as a linear combination of polarized parton densities and coefficient functions as~\cite{Shahri:2016uzl,Goto:1999by,Nath:2017ofm}
\begin{eqnarray}\label{eq:g1pxspace}
&&g_1^{\rm p} (x, Q^2) = \frac{1}{2} \sum_q e^2_q  \Delta q_{v}(x, Q^2)\otimes\nonumber\\ &&\left(1+\frac{\alpha_s(Q^2)}{2 \pi}\Delta C^{(1)}_q+\left(\frac{\alpha_s(Q^2)}{2 \pi}\right)^2\Delta C^{(2)}_{ns}\right)\nonumber\\
&&+e^2_q (\Delta q_s+\Delta \bar{q_s})(x, Q^2)\otimes \nonumber \\
&&\left(1+\frac{\alpha_s(Q^2)}{2 \pi}\Delta C^{(1)}_q+\left(\frac{\alpha_s(Q^2)}{2 \pi}\right)^2\Delta C^{(2)}_{s}\right)\nonumber\\
&&+\frac{2}{9}\Delta g(x, Q^2)\otimes
\left(\frac{\alpha_s(Q^2)}{2 \pi}\Delta C^{(1)}_g+\left(\frac{\alpha_s(Q^2)}{2 \pi}\right)^2\Delta C^{(2)}_g\right)\nonumber\\
\end{eqnarray}
Here, $\Delta q_{v}$, $\Delta q_s$ and $\Delta g$ are the polarized valance, sea and
gluon densities, respectively. The pQCD evolution kernel for PPDFs is now available at NNLO in Ref.~\cite{Moch:2014sna}.
The $\Delta C^{(1)}_q$ and $\Delta C^{(1)}_g$ are the NLO spin-dependent quark and gluon hard scattering coefficients, calculable in  pQCD~\cite{Lampe:1998eu}.

We applied the hard scattering coefficients extracted at NNLO approximation.
In this order the Wilson coefficients are different for quarks and antiquarks and we used $\Delta C^{(2)}_{ns}$ and $\Delta C^{(2)}_{s}$ \cite{Zijlstra:1993sh}.
The typical convolution in $x$ space is represented with the symbol $\otimes$.
Considering isospin symmetry, the corresponding neutron structure
functions are available.
The leading-twist deuteron structure function can be obtained from  $g_1^{\rm p}$ and $g_1^{\rm n}$ via the relation
\begin{eqnarray}\label{eq:g1dxspace}
g_{1}^{\rm \tau 2(d)}(x,Q^2) = \frac{1}{2}\{g_1^{\rm p}(x,Q^2) + g_1^{\rm n}(x,Q^2)\} \times (1 - 1.5 w_D) \,,\nonumber  \\
\end{eqnarray}
where $w_D=0.05\pm0.01$ is the probability to find the deuteron in a $D$-state~\cite{Lacombe:1981eg,Buck:1979ff,Zuilhof:1980ae}.
The leading-twist polarized structure function of $g_{2} ^{\tau 2}(x, Q^2)$ is fully determined from $g_1^{\tau 2}(x, Q^2)$ via the Wandzura and Wilczek (WW) term~\cite{Wandzura:1977qf,Flay:2016wie}:
\begin{eqnarray}\label{eq:WW}
g_{2} ^{\tau 2}(x, Q^2) & = & g_{2} ^{WW}(x, Q^2) =  \nonumber  \\
&& - g_1^{\tau 2}(x, Q^2) + \int_x^1\frac{dy}{y} g_1^{\tau 2}(y, Q^2)\,. 
\end{eqnarray}
This relation remains valid in the leading twist even though  target mass corrections are
included~\cite{Wandzura:1977qf}.

The leading-twist definition for  $g_1^{\tau 2}(x, Q^2)$ and  $g_2^{\tau 2}(x, Q^2)$ are valid in the Bjorken limit, i.e. $Q^2 \rightarrow \infty,~x=$ fixed. While, at a moderate low $Q^2$ ($\sim 1-5$ GeV$^2$) and $W^2 ($$4$ GeV$^2 <W^2<10$ GeV$^2$), TMCs and HT contributions should be considered completely in the nucleon structure functions studies.
As we have already mentioned, the most significant improvement in KTA17 analysis in comparison to Ref.~\cite{Shahri:2016uzl} is the treatment of target mass corrections and higher-twist contributions to the spin-dependent structure functions. They will be discussed in detail in the following subsections.

%
\subsection{Jacobi polynomials approach}\label{Jacobi-polynomials}
%

The method we employed in this paper is based on the Jacobi polynomials expansion of the polarized structure functions.
Practical aspects of this method including its major advantages are presented in our previous studies~\cite{Shahri:2016uzl,Khorramian:2010qa,Khorramian:2009xz,MoosaviNejad:2016ebo,Khanpour:2016uxh} and also  other literature~\cite{Ayala:2015epa,Barker:1982rv,Barker:1983iy,Krivokhizhin:1987rz,Krivokhizhin:1990ct,Chyla:1986eb,Barker:1980wu,Kataev:1997nc,Alekhin:1998df,Kataev:1999bp,Kataev:2001kk,Kataev:2005ci,Leader:1997kw}. Here, we outline a brief review of this method.
In the polynomial fitting procedure, the evolution equation is combined with the truncated series to perform a direct fit to structure functions.
According to this method, one can easily expand the polarized structure functions $x g_1^{\rm  QCD}(x,Q^2)$, in terms of the Jacobi polynomials $\Theta_{n}^{\alpha, \beta}(x)$, as follows,
\begin{equation}\label{eq:xg1QCD}
	x \, g_{1}^{\tau 2}(x, Q^2) = x^{\beta} (1 - x)^{\alpha}\ \, \sum_{n = 0}^{\rm N_{ \rm max}} a_n(Q^2) \, \Theta_n^{\alpha, \beta}(x) \,,
\end{equation}
where $\rm N_{\rm max}$ is the maximum order of the expansion. The parameters $\alpha$ and $\beta$ are Jacobi polynomials free parameters which normally fixed on their best values.
These parameters have to be chosen so as to achieve the fastest convergence of the series on the right-hand side of Eq.~\eqref{eq:xg1QCD}.

The Q$^2$ dependence of the polarized structure functions are codified in the Jacobi polynomials moments, $a_{n}(Q^{2})$. The $x$ dependence will be provided by the weight function $w^{\alpha, \beta}(x) \equiv  x^{\beta} (1 - x)^{\alpha}$ and the Jacobi polynomials $\Theta_n^{\alpha, \beta}(x)$ which can be written as,
\begin{equation}\label{eq:Jacobi}
	\Theta_n^{\alpha, \beta}(x) = \sum_{j = 0}^{n} \, c_j^{(n)}(\alpha, \beta) \, x^j \,,
\end{equation}
where the coefficients $c_j^{(n)}(\alpha, \beta)$ are combinations of Gamma functions in term of $n$, $\alpha$ and $\beta$. The above Jacobi polynomials have to satisfy the following orthogonality relation,
\begin{equation}\label{eq:orthogonality}
	\int_0^1 dx \, x^{\beta} (1 - x)^{\alpha} \, \Theta_n^{\alpha, \beta}(x) \, \Theta_{l}^{\alpha, \beta}(x) \, = \, \delta_{n, l} \,.
\end{equation}
Consequently one can obtain the Jacobi moments, $a_n(Q^2)$, using the above orthogonality relations as,
\begin{eqnarray} \label{eq:aMoment}
	a_n(Q^2) & = & \int_0^1 dx \, x g_1^{\tau 2}(x,Q^2) \, \Theta_n^{\alpha, \beta}(x)   \nonumber \\
	& = & \sum_{j = 0}^n \, c_j^{(n)}(\alpha, \beta) \, {\cal M} [xg_{1}^{\tau 2}, \, j + 2](Q^2)  \,,
\end{eqnarray}
where the Mellin transform ${\cal {M}} [x g_1^{\tau 2}, \rm N]$ is introduced as
\begin{eqnarray}\label{eq:Mellin}
	{\cal {M}} [x g_1^{\tau 2}, {\rm N}] (Q^2) \equiv  \int_0^1 dx \, x^{\rm N-2} \, xg_1^{\tau 2} (x, Q^2) \,.
\end{eqnarray}

Finally, having the QCD expressions for the Mellin moments ${\cal M} (Q^2)$, we can reconstruct the polarized structure function $x g_1^{\tau 2}(x, Q^2)$. Using the Jacobi polynomial expansion method, the $x g_1^{\tau 2}(x, Q^2)$ can be constructed as
\begin{eqnarray}\label{eq:xg1Jacobi}
x g_1^{\tau 2}(x, Q^2) & = & x^{\beta}(1 - x)^{\alpha} \, \sum_{n = 0}^{\rm N_{max}} \, \Theta_n^{\alpha, \beta}(x)    \nonumber   \\
& \times & \sum_{j = 0}^n \, c_j^{(n)}{(\alpha, \beta)} \, {\cal M}[x g_1^{\tau 2}, j + 2](Q^2) \,.
\end{eqnarray}
We have shown in our previous analyses that by setting the {$N_{\rm max}$ = 9, $\alpha$ = 3, $\beta$ = 0.5}, the optimal convergence of this expansion throughout the whole kinematic region constrained by the polarized DIS data is possible.
If $\alpha$ is allowed to vary in the fit procedure, it takes up values close to 3 with neither a change in PPDF parameter values nor a significant improvement in the $\chi^2/{\rm d.o.f}$. By contrast, in the absence of sufficiently enough data to constrain $\beta$ reasonably directly, we prefer to fix $\beta$ to the value 0.5 suggested by Regge arguments at low $x$. For the chosen $\alpha$ and $\beta$ values, the rate of convergence is adequate for all practical purposes.
The $\rm N_{max}$ can become arbitrarily large. The freedom to increase $\rm N_{max}$ can compensate for injudiciously chosen values of the constant $\alpha$ and $\beta$. However we want to deduce the expansion evolution terms and find the most practical form. To study the dependence of fit results to the value of $\rm N_{max}$, we also allow it to vary. In practice, we found that at $Q_0^2$ = 1 GeV$^2$, for $\alpha=3$ and $\beta=0.5$, no improvement is achieved by allowing polynomials expansion vary between seven and nine terms.
Inserting the Jacobi polynomial expansion of $g_1^{\tau 2}(x, Q^2)$ from Eq.~\eqref{eq:xg1Jacobi} into the WW relation Eq. \eqref{eq:WW} leads to an analytical result for the $g_2^{\tau 2}(x, Q^2)$ structure function.

%
%
\subsection{Target mass corrections and threshold problem}\label{target-mass-correction}

In the low $Q^2$ region, the nucleon mass correction cannot be neglected and the power-suppressed corrections to the structure functions can make important contributions in some kinematic regions.	
Different from the case for dynamical HT effects, the TMCs can be calculated in closed-form expression.
We follow the method suggested by Georgi and Politzer~\cite{Georgi:1976ve} in the case of the unpolarized structure function which is generalized by Blumlein and Tkabladze~ \cite{Blumlein:1998nv} for all polarized structure functions. These corrections were both presented in terms of the integer moments and Mellin inversion to $x$ space.
The explicit twist-2 expression of $g_1$ with TMCs is~\cite{Dong:2006jm,Dong:2008zg,Dong:2007iv,Dong:2007zzc,Sidorov:2006fi}
\begin{eqnarray}\label{eq:g1TMC}
&& g_1^{\rm \tau2 + TMCs}(x,Q^2)  \nonumber  \\
& = & \frac{xg_1^{\tau 2}(\xi, Q^2; {\rm M} = 0)}{\xi(1 + 4 {\rm M}^2x^2/Q^2)^{3/2}}   \nonumber  \\
& + & \frac{4{\rm M}^2x^2}{Q^2}\frac{(x + \xi)}{\xi(1 + 4 {\rm M}^2x^2/Q^2)^2}\int_{\xi}^{1}\frac{d\xi'}{\xi'} g_1^{\tau 2}(\xi', Q^2; {\rm M}=0)  \nonumber  \\
& - & \frac{4 {\rm M}^2 x^2}{Q^2}\frac{(2 - 4 {\rm M}^2x^2/Q^2)}{2(1 + 4 {\rm M}^2x^2/Q^2)^{5/2}}  \nonumber  \\
& \times & \int_{\xi}^1\frac{d \xi'}{\xi'} \int_{\xi'}^{1}\frac{d\xi''}{\xi''} g_1^{\tau 2} (\xi'', Q^2; {\rm M} = 0) \,.
\end{eqnarray}
Here, ${\rm M}$ is the nucleon mass.
Similarly, the target mass corrected structure function $g_2$  with twist-2 contribution is given by
\begin{eqnarray}\label{eq:g2TMC}
&& g_2^{\tau2+ \rm TMCs}(x,Q^2)  \nonumber  \\
& = & -\frac{xg_1^{\tau 2}(\xi, Q^2; {\rm M} = 0)}{\xi(1 + 4 {\rm M}^2x^2/Q^2)^{3/2}}  \nonumber  \\
& + & \frac{x(1 - 4 {\rm M}^2x\xi/Q^2)}{\xi(1 + 4 {\rm M}^2x^2/Q^2)^2}\int_{\xi}^{1}\frac{d\xi'}{\xi'} g_1^{\tau 2}(\xi', Q^2; {\rm M} = 0)  \nonumber  \\
& + & \frac{3}{2}\frac{4 {\rm M}^2 x^2/Q^2}{(1 + 4 {\rm M}^2 x^2/Q^2)^{5/2}}  \nonumber  \\
& \times & \int_{\xi}^1\frac{d \xi'}{\xi'}\int_{\xi'}^{1}\frac{d \xi''}{\xi''} g_1^{\tau 2}(\xi'', Q^2; {\rm M} = 0) \,,
\end{eqnarray}
where the Nachtmann variable~\cite{Nachtmann:1973mr} is given by
\begin{equation}
	\xi = \frac{2x}{1 + \sqrt{1 + 4 {\rm M}^2 x^2 / Q^2}}~.
\end{equation}
The maximum kinematic value of $\xi$ is less than unity,
which means that both the polarized and unpolarized target mass corrected leading-twist structure functions do not vanish at $x=1$. This longstanding threshold problem appears in the presence of TMCs and violates the momentum and energy conservation.
The kinematics where this problem becomes relevant are limited to the nucleon resonance region.
Many efforts were made to avoid this unphysical behavior by considering various prescriptions. It has been discussed at length in the literature \cite{Schienbein:2007gr}. These solutions are not unique~\cite{Georgi:1976ve,Piccione:1997zh,DeRujula:1976ih,Ellis:1982cd,Schienbein:2007gr,Accardi:2008pc,Sato:2016tuz}.
Accardi and Melnitchouk~\cite{Accardi:2008pc} introduced some limitations on virtuality of the struck quark to have an abrupt cutoff at $x=1$.
Where as, Georgi and Politzer~\cite{Georgi:1976ve}, Piccione and Ridolfi~\cite{Piccione:1997zh}, and also authors of~\cite{DeRujula:1976ih,Ellis:1982cd}
argued that higher-twist terms must be taken into account in the region of large $x$ to prevent the threshold problem.
Furthermore, D'Alesio {\it et al.} ~\cite{D'Alesio:2009kv} defined the maximum kinematically allowed region of $x$ by imposing the probability for hadronization as $\theta(x_{\rm{TH}}-x)$ while

\begin{equation}
x_{\rm{TH}}=\frac{Q^2}{Q^2+\mu(2M+\mu)}\ .
\label{eq:theta}
\end{equation}
Here, $\mu$ is the lowest mass particle accessible in the process of interest.
In this paper, we follow the later prescription to tame this paradox.

%
%
\subsection{Higher-twist effects }\label{Higher-twist-effect}

In addition to the pure kinematical origin TMCs, polarized structure functions in the OPE receive remarkable contributions also from HT terms.
In the range of large values of $x$, their contributions are increasingly important.
The study of HT corrections provides us direct insight into the nature of long-range dynamical multigluon exchange or parton correlation in the nucleon. Similar to the TMCs, HT terms contribute at low values of $Q^2$ and vanish at large $Q^2$.
Both $g_1$ and $g_2$ structure functions involve nonperturbative contributions from the quark and gluon correlations.
In the case of $g_1$ structure function, these correlations emerge in powers of the inverse Q$^2$ and thus are suppressed.

The $g_2(x, Q^2)$ structure function can be written as~\cite{Anselmino:1994gn}
%
\begin{equation}
g_2(x, Q^2) = g_{2} ^{\tau 2}(x, Q^2) + \bar g_2(x, Q^2)~, \label{WW}
\end{equation}
where,
\begin{equation}
\bar g_2(x, Q^2) = -\int _x ^1 \frac{\partial}{\partial y}\left[\frac{m_q}{\rm M}h_T(y,Q^2)+\zeta(y, Q^2) \right]\frac{dy}{y} \,.
\end{equation}
The function $h_T(x,Q^2)$ denotes the leading-twist transverse polarization density. Its contribution is suppressed by the ratio of the quark to nucleon masses, $\frac{m_q}{\rm M}$. The twist-3 term $\zeta(x,Q^2)$ is associated with the nonperturbative multi-parton interactions. There is no direct interpretation for these nonperturbative contributions and they can only be calculated in a model-dependent manner.

We utilized the HT parametrization form suggested by  Braun,Lautenschlager,Manashov, and Pirnay (BLMP)~\cite{Braun:2011aw}. To this end, we construct higher-twist parton distributions in a nucleon at some reference scale as,
\begin{eqnarray}\label{eq:g2HT}
	g_2^{\tau 3}(x) & = & A_{\rm {HT}} [{\rm ln}(x) + (1 - x) + \frac{1}{2}(1 - x)^2]  \nonumber  \\
	&+& (1-x)^3[B_{\rm {HT}} + C_{\rm {HT}}(1 - x) + D_{\rm {HT}} (1 - x)^2  \nonumber  \\
	&+& E_{\rm {HT}}(1 - x)^3]\,,
\end{eqnarray}
where the coefficients \{$A_{\rm HT},B_{\rm HT},C_{\rm HT},D_{\rm HT},E_{\rm HT}$\} for the proton, neutron and deuteron can be obtained by fitting to data.
Using
\begin{equation}
g_{2}^{\tau 3}(n) = \int_0^1 g_2^{\tau 3}(x) x^{n-1} \, dx \,,
\end{equation}
one can obtain the Mellin moments. The $Q^2$ dependence of the $g_{\rm 2}^{\tau 3}$ can be achieved within nonsinglet perturbative QCD evolution as
\begin{equation}
g_{\rm 2}^{\tau 3}(n, Q^2) = {\cal M}^{\rm NS}(n, Q^2) \,  g_{2}^{\tau 3}(n)\,.
\end{equation}
This method is compared with exact evolution equations for the gluon-quark-antiquark correlation in Ref.~\cite{Braun:2011aw}. Their results are
almost the same since the HT contributions are specially important in large-$x$ region. We note that by modifying the large-$x$ behavior, the small-$x$ polarized parton densities could be affected by he momentum sum rule.
Using the Jacobi polynomials technique presented in Eq.~\eqref{eq:xg1Jacobi}, one can reconstruct the twist-3 part of spin-dependent structure
functions, $x g_{\rm 2}^{\tau 3}(x, Q^2)$ ,vs its $Q^2$-dependent Mellin moments.

By the integral relation of
\begin{eqnarray}
	g_{\rm 1}^{\tau 3}(x, Q^2) = \frac{4 x^2 {\rm M}^2}{Q^2}[g_{\rm 2}^{\tau 3}(x, Q^2)  - 2\int_x^1 \frac{dy}{y} g_{\rm 2}^{\tau 3}(y, Q^2)]\ ,\nonumber \\
	\label{eq:g1HT}
\end{eqnarray}
the twist-3 part of spin-dependent structure functions, $g_{\rm 1}^{\tau 3}(x, Q^2)$, also can be obtained~\cite{Blumlein:1998nv}.
Finally, the spin-dependent structure functions considering the TMCs and  HT terms  are as follows:
\begin{eqnarray}\label{eq:g1full}
x g_{1,2}&&^{\text {Full=pQCD+TMC+HT}}(x, Q^2) = \, \nonumber \\
&&xg_{1,2}^{\rm \tau2+TMCs}(x, Q^2) + xg_{1,2}^{\tau 3}(x, Q^2)\,.
\end{eqnarray}
%
It is a particular feature of $x g_2^{\text {Full}}(x, Q^2)$ in which twist-3 term is not suppressed by inverse powers of $Q^2$ so it is equally important as its twist-2 contributions.

Here, we neglected the effect of TMCs on $\tau3$ terms, similar to JAM13 \cite{Jimenez-Delgado:2013boa}. Concerning the current level of accuracy our estimation seems reasonable.
Of course, with the new generation of data coming from 12 GeV Jefferson Lab experiments \cite{JLAB-12} our analysis should be extended to include TMCs for $\tau3$, but for now it stands reasonably well.

%
%
\section {KTA17 NNLO QCD analysis and parametrization}\label{global-PPDFs}

Motivated by the interest in studying the effects of information arising from HT effects and TMCs, we carried out the following new global analysis of PPDFs.
We present that our predictions are consistent with the results obtained in the recent studies.
In this section, we discuss the method of KTA17 analysis, including the functional form we use, the data sets considered in the analysis and the method of error calculations.
The determination of polarized PDFs uncertainties also follows the method given in this section.

%
\subsection{Parametrization}\label{Parametrization}
%

Various functional forms have been proposed so far for the polarized PDFs in pQCD analyses. Throughout our analysis, we adopt exactly the same conventions as in the TKAA16 global fit~\cite{Shahri:2016uzl}. In the present analysis,  we take into account the following parametrization at the initial scale Q$_0^2$ = 1 GeV$^2$,
\begin{equation}\label{eq:parametrizationsQ0}
x \Delta q(x, Q_0^2) = {\cal N}_{q} \, \eta_q \, x^{\alpha_q} (1 - x)^{\beta_q} \, (1 + \gamma_q x) \,,
\end{equation}
where the normalization factors, ${\cal N}_q$, can be determined as
\begin{equation}\label{eq:Norm}
\frac {1} {{\cal N}_q } = \left (1 + \gamma_q \frac{\alpha_q} {\alpha_q + \beta_q + 1} \right) \, B \left (\alpha_q, \beta_q + 1 \right) \,.
\end{equation}
%
The label of $\, \Delta q=\{\, \Delta u_v,\, \Delta d_v,\, \Delta \bar{q},\, \Delta g\}$ corresponds to the polarized up-valence,  down-valence, sea  and gluon distributions, respectively. Charm and bottom quark contributions play no role for all presently available data. $B(\alpha_q, \beta_q + 1)$ is the Euler beta function.
Considering SU(3) flavor symmetry, and due to the absence of semi-inclusive DIS (SIDIS) data in the KTA17 analysis, we attempt to fit only $\Delta \bar{q}  \equiv  \Delta \bar{u} = \Delta \bar{d} = \Delta \bar{s} = \Delta s$, while we would allow for a SU(3) symmetry breaking term by considering $\kappa$ factor such that  $ \Delta \bar{s} = \Delta s=\kappa\Delta \bar{q}$. No improvement is achieved for the specific choice of $\kappa$.

Referring to the inclusive polarized DIS World Data only, this strategy for the evolution of valence and sea quark distributions has previously been applied by Blumlein and Bottcher \cite{Blumlein:2010rn}, by the LSS group in \cite{Leader:2006xc} and also in our earlier studies \cite{Shahri:2016uzl,Monfared:2014nta,Khorramian:2010qa}.

The normalization factors, ${\cal N}_q$, are chosen such that the parameters $\eta_q$ are the first moments of $\Delta q_i(x, Q_0^2)$, as $\eta_i = \int_0^1 dx \, \Delta q_i  (x, Q_0^2)$.
The present polarized DIS data are not accurate enough to determine all the shape parameters with sufficient accuracy. Equation~(\ref{eq:parametrizationsQ0}) includes 14 free parameters in total in which we further reduce the number of free parameters in the final minimization.
The first moments of the polarized valence distribution can be described in terms of axial charges for octet baryon, $F$ and $D$ measured in hyperon and neutron $\beta$ decay. These constraints lead to the values $\eta_{u_v} = 0.928 \pm 0.014$ and $\eta_{d_v} = - 0.342 \pm 0.018$ \cite{Agashe:2014kda}. We fix two valence first moments on their central values.
The parameters $\eta_{\bar q}$ and $\eta_g$ are determined from the fit.

We find the factor $(1 + \gamma_{q} x)$ provides flexibility to achieve a good description of data, especially for the valence densities \{$\gamma_{u_v}$,$\gamma_{d_v}$\}.
The relevance of the parameters $\gamma_{\bar q}$ and $\gamma_g$ has been investigated by fixing all of them to zero and releasing them separately to test all possible combinations.
Due to the present accuracy of the polarized DIS data, no improvement is observed and we prefer to set them to zero.

The parameters \{$A_{\rm HT},B_{\rm HT},C_{\rm HT},D_{\rm HT},E_{\rm HT}$\} from Eq. (\ref{eq:g2HT}) specify the functional forms of $g_2^{\tau3}$ and consequently $g_1^{\tau3}$.
They can be extracted from a simultaneous fit to the polarized observables.

%
\subsection {Overview of data sets}
%

The core of all polarized PDFs fits comprises the DIS data obtained at the electron-proton collider and in fixed-target experiments corresponding to the proton, the neutron and heavier targets such as the deuteron.
Beside polarized DIS data, a significant amount of fixed-target SIDIS data~\cite{Alekseev:2010ub,Ackerstaff:1997ws,Alekseev:2007vi,Adeva:1997qz,Alekseev:2009ac} and the data from longitudinally polarized proton-proton ($p p$) collisions at the RHIC have only recently become available, for a limited range of momentum fractions $x$, $0.05<x<0.4$ ~\cite{Aschenauer:2015eha}.

In the KTA17 analysis, we focus on the polarized DIS data samples.
However, as only inclusive DIS data are included in the fit, it is not possible to separate quarks from antiquarks.
We include the $g_2$ structure function in the KTA17 fitting procedure, which has been traditionally neglected due to the technical difficulty in operating the required transversely polarized target.
We use all available $g_1^p$ data from E143, HERMES98, SMC, EMC, E155, HERMES06, COMPASS10 and COMPASS16 experiments ~\cite{Abe:1998wq,HERM98,Adeva:1998vv,EMCp,E155p,HERMpd,COMP1,Adolph:2015saz};  $g_1^n$ data from HERMES98, E142, E154, HERMES06, Jlab03, Jlab04, and Jlab05~\cite{HERM98,E142n,E154n,Ackerstaff:1997ws,JLABn2003,JLABn2004,JLABn2005}; and finally the $g_1^d$ data from E143, SMC, HERMES06, E155, COMPASS05, and COMPASS06~\cite{Abe:1998wq,Adeva:1998vv,HERMpd,E155d,COMP2005,COMP2006}.
The DIS data for $g_2^{p, n, d}$ from E143, E142, Jlab03, Jlab04, Jlab05, E155, Hermes12, and SMC~\cite{Abe:1998wq,E142n,JLABn2003,JLABn2004,JLABn2005,E155pdg2,hermes2012g2,SMCpg2} also are included.

These data sets are summarized in table~\ref{tab:DISdata}. The kinematic coverage, the number of data points for each given target, and the fitted normalization shifts ${\cal{N}}_i$ are also presented in this table.

To fully avoid a region of higher-twist effects, a cut in the hadronic mass $W^2$ is required. Sensitivity to the choice of cuts on $W^2$ is discussed in Ref.~\cite{Sato:2016tuz}.
It is impossible to perform such a procedure for the present data on the spin-dependent structure functions without losing too much information.
Here we want to stay inside the region of higher-twist corrections. Regarding Eq.(~\ref{eq:theta}), the maximum kinematically allowed region of $x$ is considered in our analysis. Moreover, due to the pQCD restriction, our KTA17 analysis is limited to the region of $Q^2\geq 1$ GeV$^2$.

It is already known that a reasonable choice of Q$_0^2$ is required. The DGLAP equation allows one to move in $Q^2$, provided the perturbatively calculable boundary condition.
The choice of $Q_0^2$ is typically the smallest value of $Q^2$ where the practitioner believes in pQCD. The reason is because back evolution in the DGLAP equation induces larger errors as opposed to forward evolution.
Like most of the fitting programs on the market which solve the DGLAP evolution equations in the Mellin space, the KTA17 analysis algorithm also computes the $Q^2$ evolution and extracts the structure function in $x$ space using the Jacobi polynomials approach.

%
\begin{table*}[htb]
	\caption{Summary of published polarized DIS experimental data points above $Q^2$ = 1.0 GeV$^2$ used in the KTA17 global analysis. Each experiment is given the $x$ and $Q^2$ ranges, the number
		of data points for each given target, and the fitted normalization shifts ${\cal{N}}_i$ (see the text).} \label{tab:DISdata}
	\begin{ruledtabular}
		\begin{tabular}{l c c c c c}
			\textbf{Experiment} & \textbf{Ref.} & \textbf{[$x_{\rm min}, x_{\rm max}$]}  & \textbf{Q$^2$ range {(}GeV$^2${)}}  & \textbf{Number of data points} &   \textbf{${\cal N}_n$}       \tabularnewline
			\hline\hline
			\textbf {E143(p)}   & \cite{Abe:1998wq}   & [0.031--0.749]   & 1.27--9.52 & 28 & 0.999465 \\
			\textbf{HERMES(p)} & \cite{HERM98}  & [0.028--0.66]    & 1.01--7.36 & 39 & 1.000991 \\
			\textbf{SMC(p)}    & \cite{Adeva:1998vv}    & [0.005--0.480]   & 1.30--58.0 & 12 & 0.999919 \\
			\textbf{EMC(p)}    & \cite{EMCp}     & [0.015--0.466]   & 3.50--29.5 & 10 & 1.004450 \\
			\textbf{E155}      & \cite{E155p}    & [0.015--0.750]   & 1.22--34.72 & 24 & 1.024015 \\
			\textbf{HERMES06(p)} & \cite{HERMpd} & [0.026--0.731]   & 1.12--14.29 & 51 &0.999348  \\
			\textbf{COMPASS10(p)} & \cite{COMP1} & [0.005--0.568]   & 1.10--62.10 & 15 &0.992122 \\
			\textbf{COMPASS16(p)} & \cite{Adolph:2015saz} & [0.0035--0.575]   & 1.03--96.1 & 54 &1.000009 \\
			\multicolumn{1}{c}{$\boldsymbol{g_1^p}$}       &  &  &  &  \textbf{233}  &   \\
			\textbf{E143(d)}  &\cite{Abe:1998wq}    & [0.031--0.749]   & 1.27--9.52    & 28 &  0.999005 \\
			\textbf{E155(d)}  &\cite{E155d}     & [0.015--0.750]   & 1.22--34.79   & 24 & 1.000036 \\
			\textbf{SMC(d)}   &\cite{Adeva:1998vv}     & [0.005--0.479]   & 1.30--54.80   & 12 &0.999992 \\
			\textbf{HERMES06(d)} & \cite{HERMpd}& []0.026--0.731]   & 1.12--14.29   & 51 & 0.998055   \\
			\textbf{COMPASS05(d)}& \cite{COMP2005}& [0.0051--0.4740] & 1.18--47.5   & 11 & 0.996973  \\
			\textbf{COMPASS06(d)}& \cite{COMP2006}& [0.0046--0.566] & 1.10--55.3    & 15 & 0.999949  \\
			\multicolumn{1}{c}{ $\boldsymbol{g_1^d}$}      &  &  & & \textbf{141} &     \\
			\textbf{E142(n)}   &\cite{E142n}    & [0.035--0.466]   & 1.10--5.50    & 8 & 0.998994 \\
			\textbf{HERMES(n)} &\cite{HERM98}   & [0.033--0.464]   & 1.22--5.25    & 9 & 0.999968 \\
			\textbf{E154(n)}   &\cite{E154n}    & [0.017--0.564]   & 1.20--15.00   & 17 & 0.999608 \\
			\textbf{HERMES06(n)} &\cite{Ackerstaff:1997ws}  &  [0.026--0.731]  & 1.12--14.29   & 51 & 1.000118 \\
			\textbf{Jlab03(n)}&\cite{JLABn2003} & ]0.14--0.22]     & 1.09--1.46    & 4 & 0.999728 \\
			\textbf{Jlab04(n)}&\cite{JLABn2004} & [0.33--0.60]      & 2.71--4.8     & 3 & 0.900000 \\
			\textbf{Jlab05(n)}&\cite{JLABn2005} & [0.19--0.20]     &1.13--1.34     & 2 & 1.030771 \\
			\multicolumn{1}{c}{$\boldsymbol{g_1^n}$}     &     &  & & \textbf{94} &   \\
			\textbf{E143(p)}    & \cite{Abe:1998wq}   & [0.038--0.595]   & 1.49--8.85    & 12&1.000545 \\
			\textbf{E155(p)}   &\cite{E155pdg2}  &[0.038--0.780]    & 1.1--8.4      & 8 &0.997275 \\
			\textbf{Hermes12(p)}&\cite{hermes2012g2} &[0.039--0.678]&1.09--10.35   & 20&0.998658 \\
			\textbf{SMC(p)}      &\cite{SMCpg2} & [0.010--0.378]    & 1.36--17.07   & 6 &1.000002 \\
			\multicolumn{1}{c}{$\boldsymbol{g_2^p}$}  &  &  & & \textbf{46} &    \\
			\textbf{E143(d)}     &\cite{Abe:1998wq} & [0.038--0.595]   & 1.49--8.86    & 12 &0.999985 \\
			\textbf{E155(d)}    &\cite{E155pdg2}& [0.038--0.780]    & 1.1--8.2      & 8 &1.002186 \\
			\multicolumn{1}{c}{$\boldsymbol{g_2^d}$}  &  &  & & \textbf{20} &   \\
			\textbf{E143(n)}    &\cite{Abe:1998wq}  & [0.038--0.595]   & 1.49--8.86    & 12&0.999984 \\
			\textbf{E155(n)}    &\cite{E155pdg2}&[0.038--0.780]    &1.1--8.8       & 8 &1.002422 \\
			\textbf{E142(n)}    &\cite{E142n}   &[0.036--0.466]    &1.1--5.5       & 8 &0.999981 \\
			\textbf{Jlab03(n)}  &\cite{JLABn2003}&[0.14--0.22]     & 1.09--1.46    & 4 &1.004973 \\
			\textbf{Jlab04(n)}  &\cite{JLABn2004}&[0.33--0.60]     & 2.71--4.83    & 3 &1.062181 \\
			\textbf{Jlab05(n)}  &\cite{JLABn2005}&[0.19--0.20]     & 1.13--1.34    & 2 &0.979031 \\
			\multicolumn{1}{c}{$\boldsymbol{g_2^n}$}  &  &  & &\textbf{37} &   \\  \hline
			\multicolumn{1}{c}{\textbf{ Total}}&\multicolumn{5}{c}{~~~~~~~~~~~~~~~~~~~~~~~~~~~~~~~~~~~~~~~~~~~~~~~~~~~~~~~~~~~\textbf{571}}
			\\
		\end{tabular}
	\end{ruledtabular}
\end{table*}
%

%
\subsection{ $\chi^2$ minimization }
%

To determine the best fit at NNLO, we need to minimize the $\chi^2_{\rm global}$ function with the free unknown PPDF parameters together with $\Lambda_{\rm QCD}$.
$\chi_{\rm global}^2(\rm p)$ quantifies the goodness of fit to the data for a set of independent parameters $\rm p$ that specifies the polarized PDFs at Q$_0^2$ = 1 GeV$^2$.
This function is expressed as follows,
%
\begin{equation}\label{eq:chi2}
\chi_{\rm global}^2 ({\rm p}) = \sum_{n=1}^{N_{\rm exp}} w_n \chi_n^2\,,
\end{equation}
while ${\rm w}_n$ is a weight factor for the nth experiment and
\begin{equation}\label{eq:chi2global}
\chi_n^2 (\rm p) = \left( \frac{1 -{\cal N}_n }{\Delta{\cal N}_n}\right)^2 + \sum_{i=1}^{N_n^{\rm data}} \left(\frac{{\cal N}_n  \, g_{(1,2), i}^{\rm Exp} - g_{(1,2), i}^{\rm Theory} (p) }{{\cal N}_n \, \Delta g_{(1,2), i}^{\rm Exp}} \right)^2\,.
\end{equation}
%
The minimization of the above $\chi_{\mathrm{\rm global}}^2 (\rm p)$ function is done using the CERN program library MINUIT~\cite{James:1994vla}.
In the above equation, the main contribution comes from the difference between the model and the DIS data within the statistical precision.
In the $\chi_n^2$ function, $g^{\rm Exp}$, $\Delta g^{\rm Exp}$, and $g^{\rm Theory}$ indicate the experimental measurement, the experimental uncertainty (statistical and systematic combined in quadrature) and the theoretical value for the ith data point, respectively.
${\cal N}_n$ is overall normalization factors for the data of experiment $n$ and the ${\Delta{\cal N}_n}$ is the experimental normalization uncertainty.
We allow for a relative normalization factor ${\cal N}_n$ between different experimental data sets within uncertainties ${\Delta{\cal N}_n}$ quoted
by the experiments. The normalization factors appear as free parameters in the fit. They are determined simultaneously with the parameters of the functional forms at prefitting procedure and fixed at their best values.

\subsection{ PPDFs uncertainties}
%

A robust treatment of uncertainty is desirable throughout full NNLO analysis. In this section, we briefly review the method in which we use to extract the polarized PDF uncertainties.
The methodologies for the estimation of uncertainties are essential for understanding of the accuracy of collider predictions, both for the precision
measurements and for the new physics searches.
Three approaches are available to propagate the statistical precision of the experimental data to the fit results. They are based on the diagonalization of the Hessian error matrix, the Lagrange multiplier and the Monte Carlo sampling of parton distributions~\cite{Martin:2009iq}. The Hessian and Monte Carlo techniques are the most commonly used methods.
The adequacy of parametrization Eq.~\eqref{eq:parametrizationsQ0} at the reference scale of Q$_0^2$ = 1 GeV$^2$ for given $N_{\rm max}$, $\alpha$ and $\beta$ is investigated by the Hessian matrix method which is fully discussed in  Refs.~\cite{Hou:2016sho,Khanpour:2016pph,Pumplin:2001ct,Martin:2002aw,Martin:2009iq,Shoeibi:2017lrl}.
In the Hessian method, the uncertainty on a polarized PDF, $\Delta q(x)$, can be obtained from linear error propagation,
\begin{eqnarray}\label{eq:uncertainties}
&&[\Delta q (x)]^2 = \Delta \chi^2_{\rm global} \times \, \nonumber \\
&& \big [ \sum_i (\frac{\partial \Delta q(x, {\hat a})}{\partial a_i})^2 \,  C_{i i}  + \sum_{i \neq j} ( \frac{\partial \Delta q(x, {\hat a})}{\partial a_i} \frac{\partial \Delta q(x, {\hat a})}{\partial a_j} ) \,  C_{i j}  \big ],  \nonumber \\
\end{eqnarray}
where $a_i$ ($i$ = 1, 2, ..., N) denotes to the free parameters for each distribution presented in Eq.~\eqref{eq:parametrizationsQ0}. N is
the number of optimized parameters and ${\hat a}_i$ is the optimized parameter.$C \equiv H_{i, j}^{-1}$ are the elements of the covariance matrix (or error matrix) determined in the QCD analysis at the scale $Q^2_0$. The  $T = \Delta \chi^2_{\rm global}$ is the tolerance for the required confidence region (C.L.).

In order to compare the uncertainties of polarized PDFs obtained from the present  KTA17 analysis with those obtained by other groups, we follow the standard parameter-fitting criterion considering $T = \Delta \chi^2_{\rm global}$ = 1 for 68\% (1-$\sigma$) \text{C.L.}.
It is worth noting that, the various groups have different approaches to obtain \text{C.L.} criteria for the value of $\chi^2$ in the goodness-of-fit test~\cite{Accardi:2016qay,Alekhin:2013nda,Dulat:2015mca,Abramowicz:2015mha,Jimenez-Delgado:2014twa,Harland-Lang:2014zoa}.
The difference originates from the quality of the experimental data sets. One approach is to fit to a very wide set of data (a tolerance criterion for $\Delta \chi^2$ should be introduced), while the other one rejects inconsistent data sets ( $\Delta \chi^2$ = 1).

It should also be stressed that, in the process of the analysis of NNPDF~\cite{Nocera:2014gqa,Ball:2013tyh,Ball:2013lla} or JAM~\cite{Sato:2016tuz} groups a Monte Carlo method is used to estimate the PDF uncertainty. This method allows a more robust extraction of polarized PDFs with statistically rigorous uncertainties.

In Sec.~\ref{Comparison-with-others}, we discuss the polarized PDF uncertainties in the kinematic region covered by the polarized inclusive DIS data used in this analysis.

%
%
\section{Discussion of fit results}\label{Results-of-PPDFs}

To distinguish the effect of TMCs and HT contribution, we perform three analyses as the pQCD, `pQCD+TMC', and `pQCD+TMC+HT' scenarios.
In the pQCD analysis, we only consider the leading-twist contribution of $g_1$ and $g_2$ structure functions, Eqs.~(\ref{eq:g1pxspace}, \ref{eq:g1dxspace}, and \ref{eq:WW}), while in the pQCD+TMC analysis, the TMCs are included, Eqs.~(\ref{eq:g1TMC} and \ref{eq:g2TMC}). The pQCD+TMC+HT analysis, which we referred to as KTA17, represents the effect of both TMC and HT contributions, Eq.~\eqref{eq:g1full}.
As discussed earlier, the parameters \{$\eta_{u_v},\eta_{d_v},\gamma_{\bar q},\gamma_g$\} from Eq.~\eqref{eq:parametrizationsQ0} are frozen in the first minimization step.
We start to minimize the $\chi_{\rm global}^2$ value with the $12$ unknown fit parameters of Eq.~\eqref{eq:parametrizationsQ0} and $15$ HT parameters of Eq.~(\ref{eq:g2HT}) plus an undetermined coupling constant.
Then, in the final minimization step, we fix \{$\gamma_{u_v},\gamma_{d_v},\beta_{\bar q},\beta_g$\} together with \{$A_{\rm HT},B_{\rm HT},C_{\rm HT},D_{\rm HT},E_{\rm HT}$\} for the proton, neutron, and deuteron on their optimal values determined on prefitting scenario.
As previously mentioned in Sec. \ref{Parametrization}, due to the lack of precise data, some of the parameters have to be fixed after an initial minimization step to their best values.
KTA17 results are demonstrated in Tables~\ref{TablePPDfs} and \ref{tw-3-parameters}, while parameters marked with $^*$ are fixed.
Accordingly, there are nine unknown parameters including the strong coupling constant  which provide enough flexibility to have a reliable fit.

\begin{table*}[!htb]
	\caption{  \label{TablePPDfs} Obtained parameter values and their statistical
		errors at the input scale Q$_0^2$ = 1 GeV$^2$ determined from pQCD, pQCD+TMC and pQCD+TMC+HT analyses in NNLO approximation. Those marked with ($^*$) are fixed.}
	{\setlength{\extrarowheight}{2pt}%
	\begin{tabular}{|c|c|c|c|c|c|}
		\hline  \hline
		\multicolumn{2}{|c|}  {\textbf{Parameters}}&   \textbf{pQCD}  &  \textbf{pQCD+TMC} &  \textbf{pQCD+TMC+HT (KTA17)}
		\\
		\hline\hline
		$\delta u_v$ & $\eta_{u_v} $ & $~0.928^*~$ & $~0.928^*~$   &$~0.928^*~$  \\
		& $\alpha_{u_v}$ &$0.222 \pm 0.019$ &  $0.571 \pm 0.010$  &$0.450 \pm 0.027$   \\
		& $\beta_{u_v}$ &$2.827 \pm 0.041$ &  $3.155 \pm 0.040$  & $2.971 \pm 0.102$   \\
		& $\gamma_{u_v}$ &$39.826^*$ & $6.694^*$  &$12.580^*$   \\  \hline
		$\delta {d_v}$ & $\eta_{d_{v}} $ & $-0.342^*$& $-0.342^*$ & $-0.342^*$    \\
		& $\alpha_{d_v}$ &  $0.132 \pm 0.562$& $0.160 \pm 0.477$ & $0.215 \pm 0.051$   \\
		& $\beta_{d_v}$ &  $2.856 \pm 0.267$& $3.069 \pm 0.442$ &$2.943 \pm 0.235$   \\
		& $\gamma_{d_v}$ &  $~37.918^*~$& $10.659^*$ & $~8.224^*~$   \\ \hline
		$~\delta_{\bar{q}}$ & $~\eta_{\bar{q}}$ &   $-0.098 \pm 0.004$& $-0.095 \pm 0.009$  &$ -0.099 \pm 0.002$  \\
		& $\alpha_{\bar{q}}$ &  $0.274 \pm 0.027$ & $0.350 \pm 0.043$&  $0.271 \pm 0.048$   \\
		& $\beta_{\bar{q}}$ &   $7.964^*$  &$2.606^*$ & $2.556^*$   \\
		& $\gamma_{\bar{q}}$ & $~0.0^*~$& $~0.0^*~$ & $~0.0^*~$  \\     \hline
		$\delta g$ & $\eta_g $  & $0.165 \pm 0.014$ &$0.108 \pm 0.012$& $0.111 \pm 0.046$  \\
		& $\alpha_g$ &  $13.015 \pm 0.828$   & $11.391 \pm 0.881$  &$9.090 \pm 1.175$   \\
		& $\beta_g$ &   $50.637^*$  &  $48.151^*$   &$42.586^*$   \\
		& $\gamma_g$ & $~0.0^*~$& $~0.0^*~$        & $~0.0^*~$  \\  \hline
		\hline
		\multicolumn{2}{|c|}{ $\alpha_{s} (Q_0^2)$} &{$0.4355 \pm 0.0081$} & {$0.3682 \pm 0.0093$}& {$0.3458 \pm 0.0166$}   \\
		\hline\multicolumn{2}{|c|}{ $\alpha_{s} (M_Z^2)$} &{$0.1212\pm0.0005$} & {$0.1173 \pm 0.0009$}& {$0.1157 \pm 0.028$}   \\
		\hline \multicolumn{2}{|c|}{$\chi^2/{\rm d.o.f}$} &   {$883.92/562 = 1.584$} & {$526.67/562 = 0.937$} &{$501.13/562 = 0.891$}   \\
		\hline\hline
	\end{tabular}}
\end{table*}

\begin{table}[!htb]
	\caption{ Parameter values for the coefficients of the twist-3 corrections at Q$_0^2 = 1$ GeV$^2$ obtained at NNLO approximation in the pQCD+TMC+HT analysis. Due to the large errors of the data, all the HT parameters are fixed after an initial minimization to their best values. \label{tw-3-parameters}}
	\begin{ruledtabular}
		{\setlength{\extrarowheight}{2pt}%
		\begin{tabular}{lccccc}
			& ${\rm A_{HT}}$  & ${\rm B_{HT}}$  &   ${\rm C_{HT}}$  &  ${\rm D_{HT}}$  & ${\rm E_{HT}}$    \\   \hline\hline
			$g_{2,p}^{\tau 3}$  & $0.0055$ &$0.2667$ & $0.2417$  &  $-1.4453$ &  $0.8861$    \\
			$g_{2,n}^{\tau 3}$  & $0.0099$ &$0.2196$ & $-0.3936$ &  $0.1472$  &  $-0.0100$    \\
			$g_{2,d}^{\tau 3}$  & $0.7726$ &$1.0729$ & $-1.6477$ &  $0.4758$  &  $1.4223$   \\
		\end{tabular}}
	\end{ruledtabular}
\end{table}

The $\chi^2/{\rm d.o.f.}$ of the pQCD+TMC+HT analysis is lower than both the pQCD+TMC and pQCD scenarios, indicating the significance of small- $Q^2$ corrections. Large $\chi^2/{\rm d.o.f.}$ of pQCD fit analysis confirms our theoretical  assumption in which the leading-twist part should be accompanied by both TMCs and HT terms.
As represented in Table~\ref{TablePPDfs}, all the extracted strong coupling constants at $Z$ mass are consistent with the world average value of $0.1185 \pm 0.0006$~\cite{Agashe:2014kda,Olive:2016xmw}. The $\alpha_{s} (M_Z^2)$ based on `pQCD+TMC+HT' scenario, receives 2.07\% (0.68\%) corrections including TMC+HT (HT) effects.

%
%
\subsection{NNLO polarized PDFs}\label{NNLO-PPDFs}

\begin{figure}[!htb]
	\includegraphics[clip,width=0.45\textwidth]{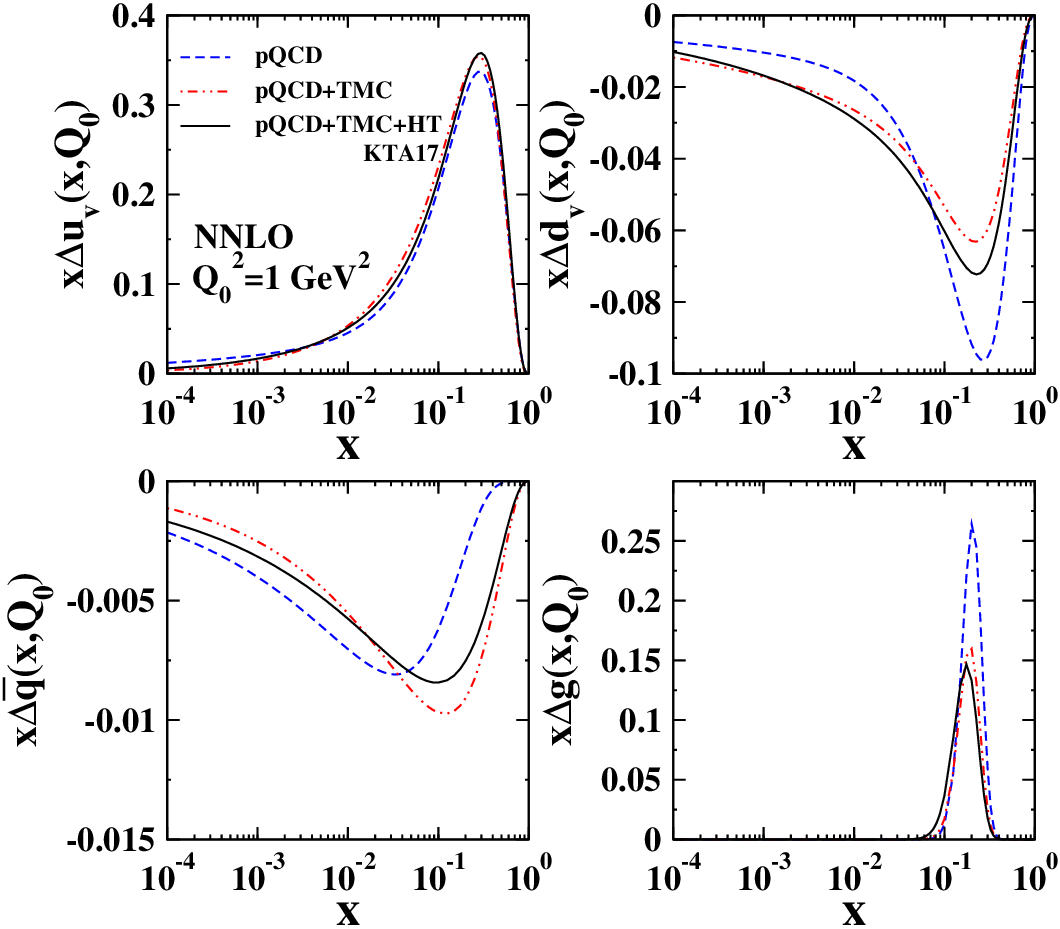}
	\begin{center}
		\caption{{\small (Color online) Our results for the polarized PDFs at Q$_0^2$= 1 GeV$^2$
				as a function of $x$ in NNLO approximation plotted as a solid curve.
				Also shown are our NNLO PPDFs based on pQCD fit (dashed) and pQCD+TMC fit (dashed-dotted-dotted).  \label{fig:partonqcdhttmc}}}
	\end{center}
\end{figure}

The effect of considering TMCs and HT terms on the   KTA17 PPDFs, $x\Delta u_v (x, Q^2)$, $x\Delta d_v (x, Q^2)$, $x\Delta \bar{q} (x, Q^2)$ and $x\Delta g (x, Q^2)$,
is individually illustrated in Fig.~\ref{fig:partonqcdhttmc}.
Including TMCs imposes significant effects on the whole $x$ region of sea quark density while valence and gluon densities are mainly affected in the large-$x$ region.

Comparing  the	pQCD+TMC and pQCD+TMC+HT curves we observe that all densities are practically identical in the small-$x$ region (except for the sea quark density); little differences appear in their peak region behavior.

\begin{figure}[!htb]
	\vspace*{0.5cm}
	\includegraphics[clip,width=0.45\textwidth]{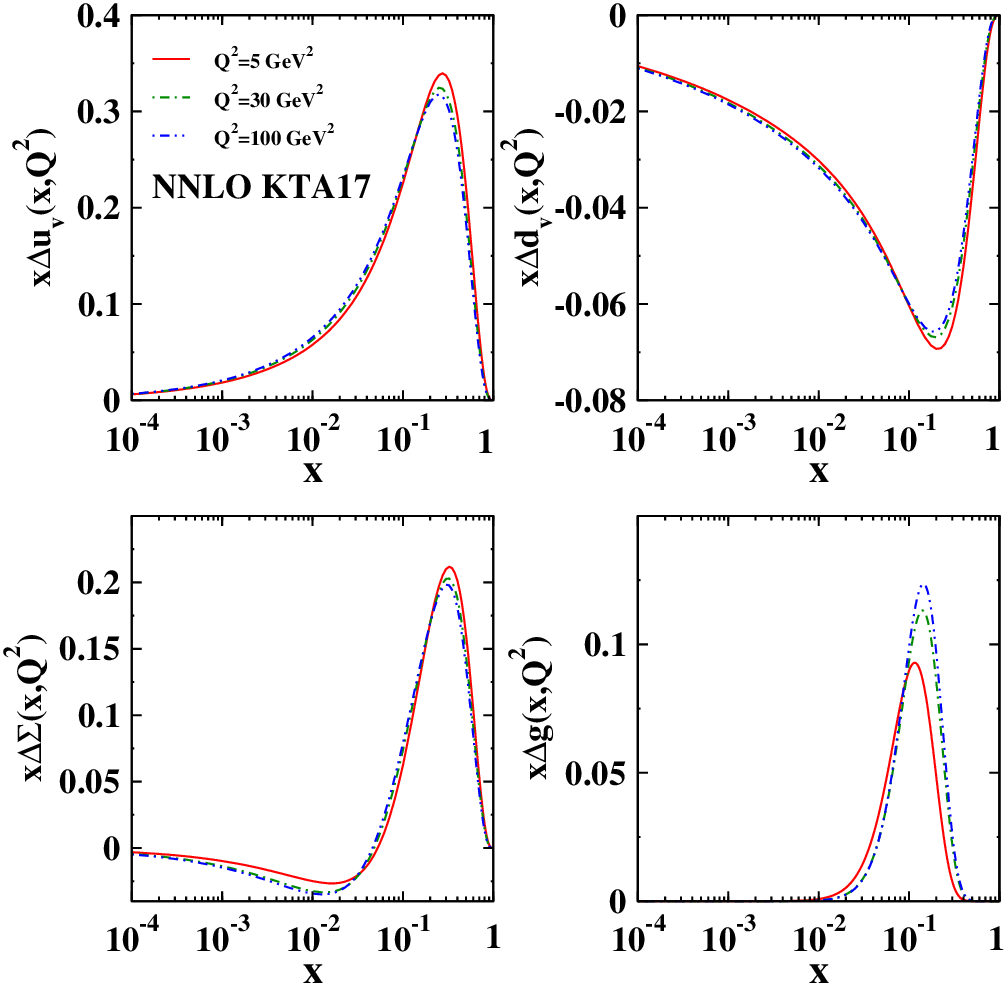}
	\begin{center}
		\caption{ \small (Color online) The  KTA17 polarized parton distributions as a function of $x$ and for some selected value of Q$^2$ = 5, 30, 100 GeV$^2$. \label{fig:partonQ} }
	\end{center}
\end{figure}

Figure~\ref{fig:partonQ} illustrates the evolution of  KTA17 polarized parton distributions for a selection of $Q^2$ values of 5, 30, and 100 GeV$^2$.
We observe that the evolution in all the distributions, except the gluon density, tends to flatten out the peak for increasing $Q^2$, While the gluon distribution increases in the large kinematic region of $x$.
	
%
%
\subsection{ Polarized PDFs comparison }\label{Comparison-with-others}

\begin{figure}[!htb]
	\vspace*{0.5cm}
	\includegraphics[clip,width=0.5\textwidth]{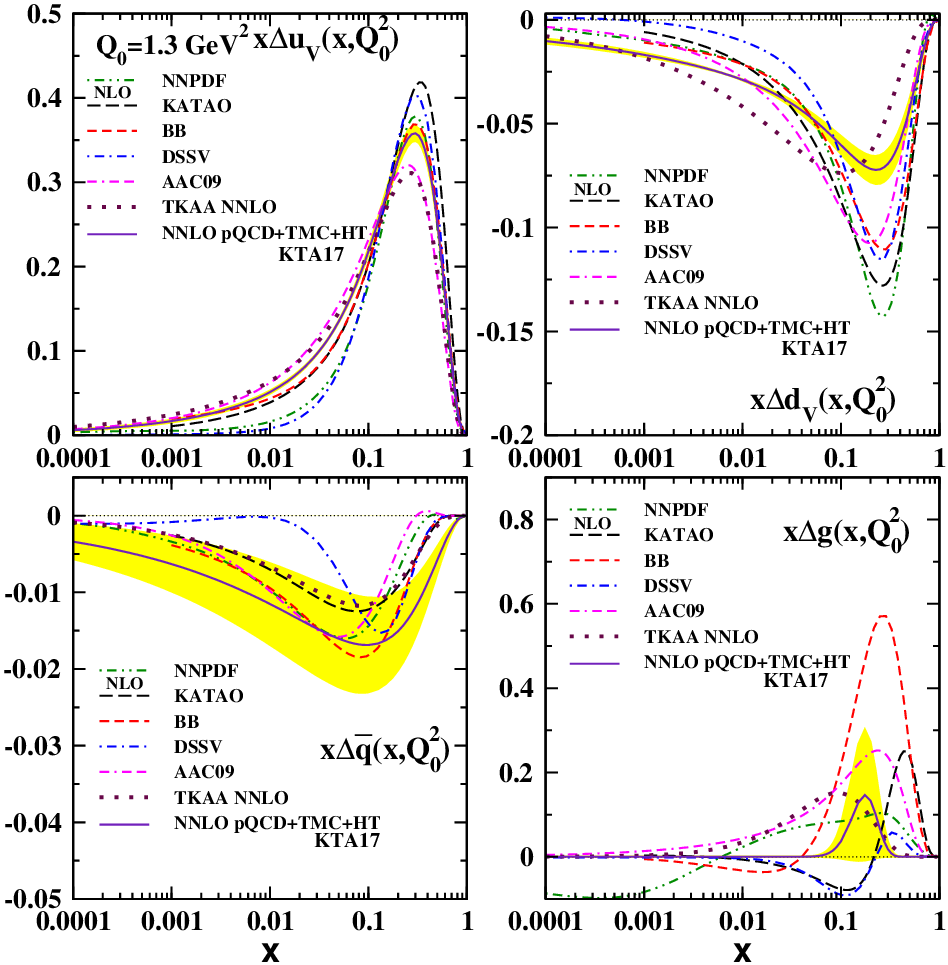}
	\begin{center}
		\caption{{\small (Color online)  KTA17  results for the polarized PDFs at Q$_0^2=$ 1 GeV$^2$
				as a function of $x$ in NNLO approximation plotted as a solid curve along with their $\Delta \chi^2 = 1$
				uncertainty bands computed with the Hessian approach, as described in the text.
				We also show the result
				obtained in earlier global analyses of NNPDF(dashed-dotted-dotted)~\cite{Nocera:2014gqa}, KATAO (long dashed)~\cite{Khorramian:2010qa}, BB (dashed)~\cite{Blumlein:2010rn}, DSSV~(dashed-dotted)
				\cite{deFlorian:2008mr}, AAC09 (dashed-dashed-dotted)~\cite{Hirai:2008aj} in NLO approximation and TKAA16 (dotted)~\cite{Shahri:2016uzl} in the NNLO approximation. \label{fig:partonQ0polarizeTMC}}}
	\end{center}
\end{figure}

We present KTA17  PPDFs along with the corresponding uncertainty bounds as a function of $x$ at Q$_0^2=$ 1 GeV$^2$ in Fig.~\ref{fig:partonQ0polarizeTMC}.
Various parameterizations of NNPDF~\cite{Nocera:2014gqa}
KATAO~\cite{Khorramian:2010qa}, BB10~\cite{Blumlein:2010rn}, DSSV09~\cite{deFlorian:2008mr},
AAC09~\cite{Hirai:2008aj}, AKS14~\cite{Arbabifar:2013tma},  LSS06~\cite{Leader:2006xc} and THK14~\cite{Monfared:2014nta} at the NLO approximation, and TKAA16~\cite{Shahri:2016uzl} at the NNLO approximation are illustrated for comparison.
In the polarized PDF sets (NNPDF, LSS and DSSV) which include SIDIS and/or W boson production in polarized $pp$ collisions, $\Delta\bar u$ is different from $\Delta\bar d$, which are in turn different from $\frac{1}{2} (\Delta s + \Delta\bar s)$. So we considered $\Delta\bar q=\frac{1}{2} (\Delta\bar u + \Delta\bar d)$ in Fig.~\ref{fig:partonQ0polarizeTMC}.

Our uncertainty estimation is based on the Hessian methods, for a tolerance of $\Delta \chi^2 = 1$. The $x\Delta u_v$ and $x\Delta d_v$ polarized PDFs are the best determined distributions from the inclusive polarized DIS data, with relatively smaller uncertainty bands for the $x\Delta u_v$ distribution.
As one can see, our $x\Delta u_v$ is relatively compatible with other results while the $x\Delta d_v$, $x\Delta \bar q$ and $x\Delta g$ densities are treated differently.
For the extrapolated regions, $x < 10^{-3}$ and $x > 0.8$, where the PPDFs are not directly constrained by the data, all valence distributions are treated the same.

The polarized gluon distribution is the most complicated case for PPDF uncertainties and parameterizations. Results for $x\Delta g$ from the various fits are usually quite spread. As illustrated in Fig.~\ref{fig:partonQ0polarizeTMC}, the difficulty in constraining the polarized gluon distribution is clearly revealed through the spread of $x\Delta g$ from various global PPDF parametrizations.
All the gluon distributions are positive at whole $x$ range, except for the KATAO, DSSV and NNPDF which indicate a sign change. The NNPF gluon density is treated differently in the small-$x$ region.
The $x\Delta g$ distributions based on different group analyses tend to zero less quickly than the KTA17 result.

Large differences are visible over the whole $x$ range for the sea quark distribution. This distribution is actually not well constrained by the present polarized DIS data.
It should be stressed again that, in both of our NNLO analyses, we used the inclusive DIS data to constrain polarized parton distributions.
In contrast, in the fits of the LSS and DSSV collaborations,(SIDIS) data which are sensitive to the quark flavours are included. The quark-antiquark separation is achieved in NNPDF thanks to W boson production in polarized $pp$ collisions.

\begin{figure}[!htb]
		\includegraphics[clip,width=0.5\textwidth]{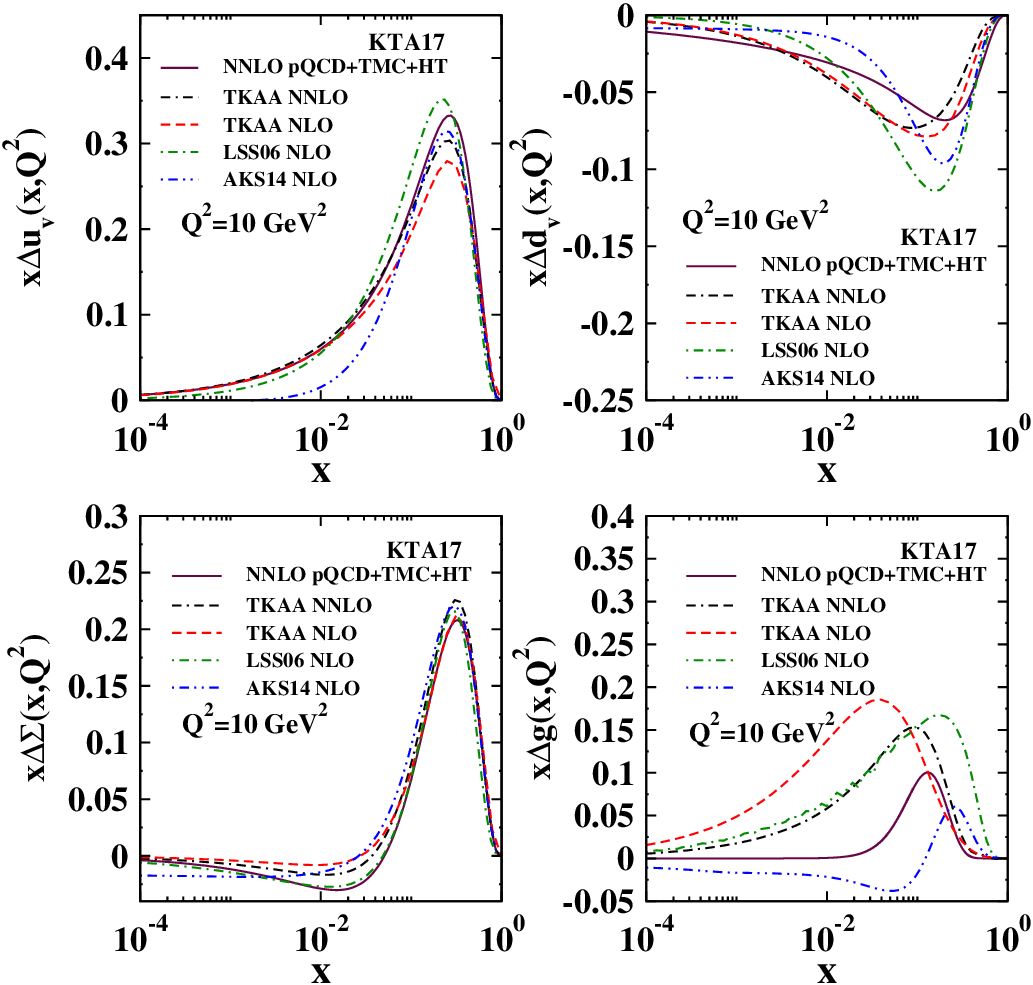}
	\vspace{-0.5cm}
	\begin{center}
		\caption{{\small (Color online) KTA17 results for the polarized PDFs at Q$^2=$ 10 GeV$^2$
				as a function of $x$ in NNLO approximation plotted as solid curves.
				Also shown are the recent results of TKAA16 (dotted)~\cite{Shahri:2016uzl} in both NLO and NNLO approximations,  AKS14~\cite{Arbabifar:2013tma} and LSS06~\cite{Leader:2006xc} analysis. \label{fig:partonNNLOQ10}}}
	\end{center}
\end{figure}

A detailed PPDF comparison is presented in Fig.~\ref{fig:partonNNLOQ10}, in which we plotted KTA17 together with those of TKAA16 (NLO and NNLO), AKS14 and LSS06 at Q$^2=$ 10 GeV$^2$ as a function of $x$. Similar to previous comparisons, gluon density remains puzzling. The gluons from all PPDF sets are positive except for the  AKS14 group which shows a sign change.
The  $x\Delta u_v$ and $x\Delta d_v$  polarized PDFs of the TKAA16 (NLO and NNLO),  AKS14  and LSS06 are qualitatively similar, though for LSS06 $x\Delta u_v$ are typically larger at medium $x$.

%

%
\subsection{Polarized structure function comparison }
%

Several efforts to study the nucleon structure have been developed, aiming to predict the polarized PDFs behavior at small and large $x$.
In order to investigate the precision of the obtained polarized PDFs and also to test whether the DIS data favor or unfavor them, a detailed comparison of the extracted structure functions and the available polarized DIS data is required.

It should be stressed that much more numerous and more accurate data at both small and large $x$ are required  to discriminate among different groups analyses.
We will return to this subject in Sec.~\ref{high-precision-era}, considering an ongoing planned and proposed high-energy polarized collider.

\begin{figure}[!htb]
	\includegraphics[clip,width=0.5\textwidth]{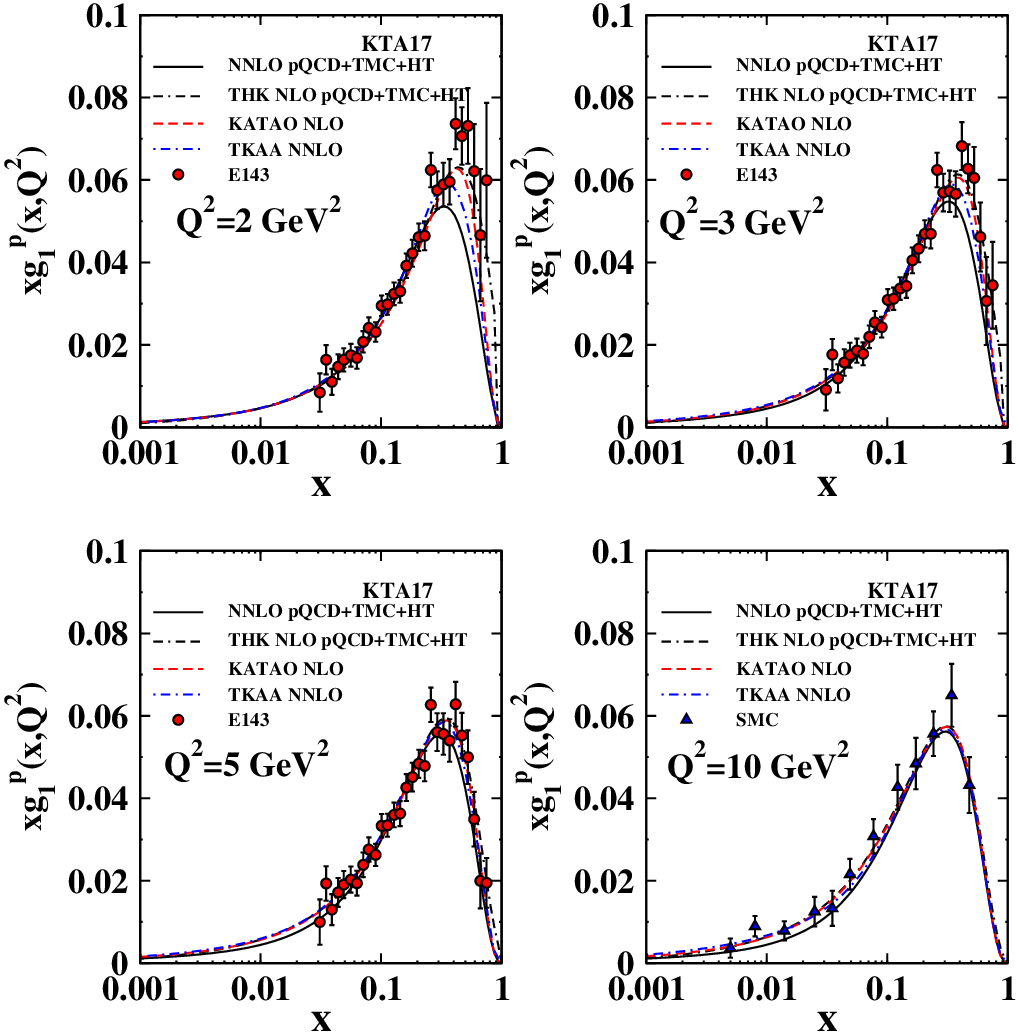}
	\begin{center}
		\caption{\small (Color online) The spin-dependent proton structure functions as a function of $x$ and Q$^2$. KTA17 (solid curve) is compared with THK (dashed-dashed-dotted) \cite{Monfared:2014nta}, KATAO (dashed) \cite{Khorramian:2010qa}, TKAA16 (dashed dotted) \cite{Shahri:2016uzl}. \label{fig:xg1p}}
	\end{center}
\end{figure}

\begin{figure}[!htb]
	\vspace*{0.5cm}
	\includegraphics[clip,width=0.5\textwidth]{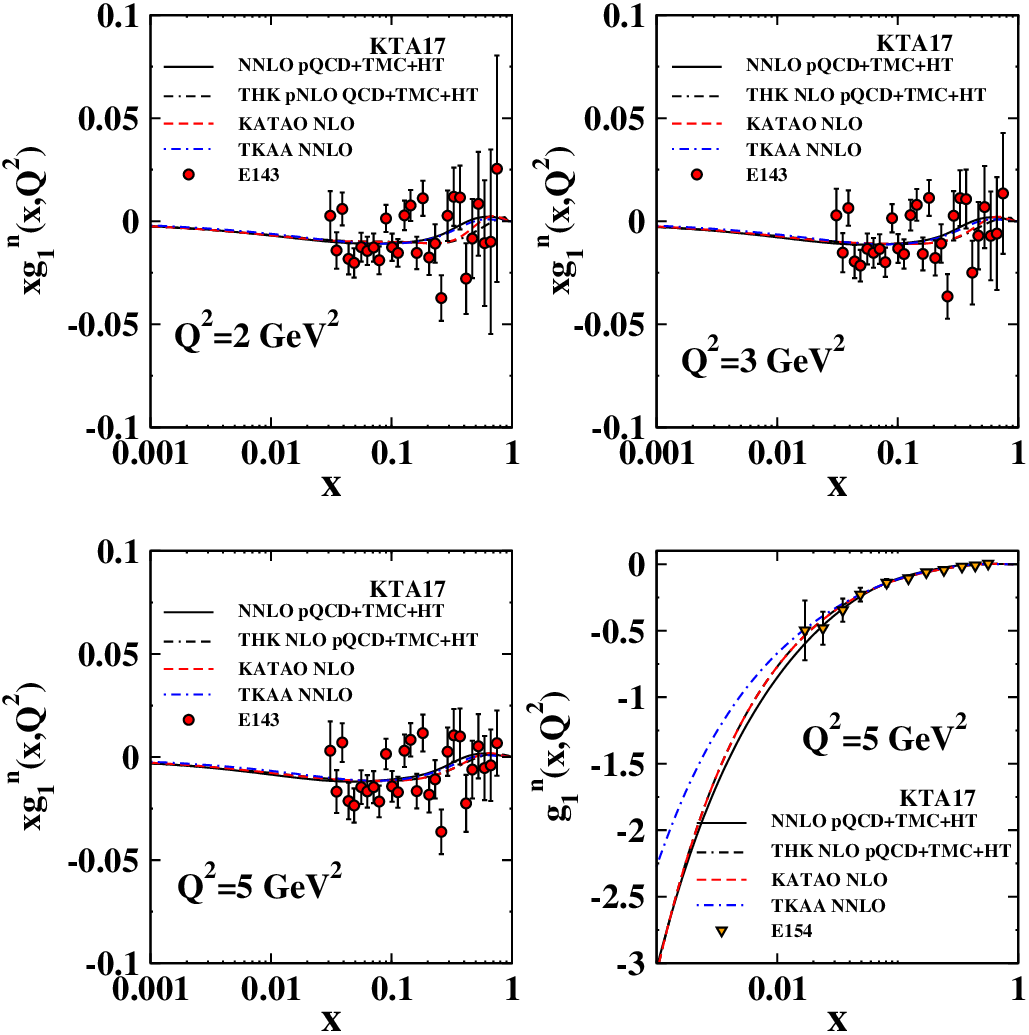}
	\begin{center}
		\caption{\small (Color online) The spin-dependent neutron structure functions as a function of $x$ and Q$^2$. KTA17 (solid curve) is compared with THK (dashed-dashed-dotted) \cite{Monfared:2014nta}, KATAO (dashed) \cite{Khorramian:2010qa}, TKAA16 (dashed dotted) \cite{Shahri:2016uzl} \label{fig:xg1n}}
	\end{center}
\end{figure}

\begin{figure}[!htb]
	\vspace*{0.5cm}
	\includegraphics[clip,width=0.4\textwidth]{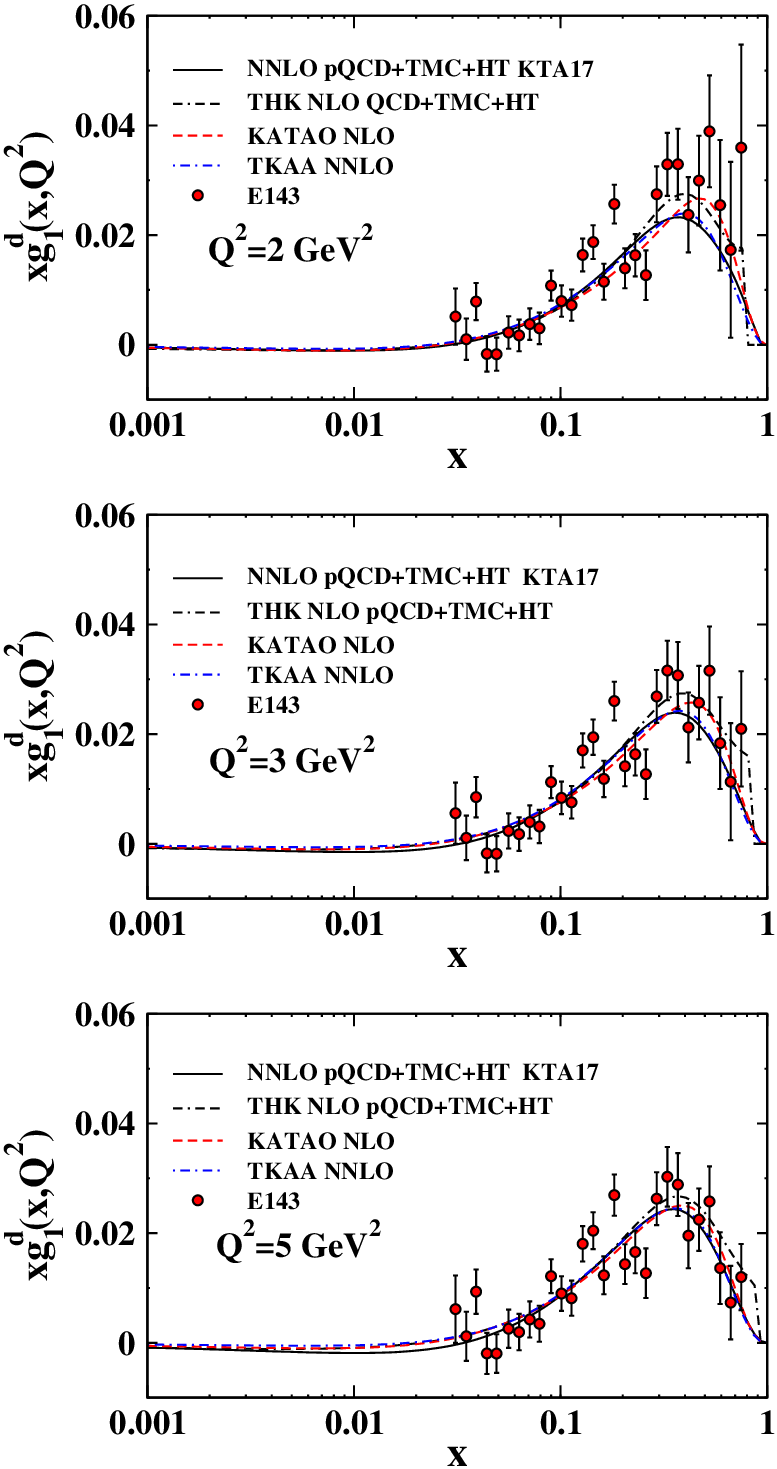}
	\begin{center}
		\caption{\small (Color online) The spin-dependent deuteron structure functions as a function of $x$ and Q$^2$. KTA17  (solid curve) is compared with THK (dashed-dashed-dotted) \cite{Monfared:2014nta}, KATAO (dashed)~\cite{Khorramian:2010qa}, TKAA16 (dashed dotted) ~\cite{Shahri:2016uzl} \label{fig:xg1d}}
	\end{center}
\end{figure}

\begin{figure}[!htb]
	\vspace*{0.5cm}
	\includegraphics[clip,width=0.45\textwidth]{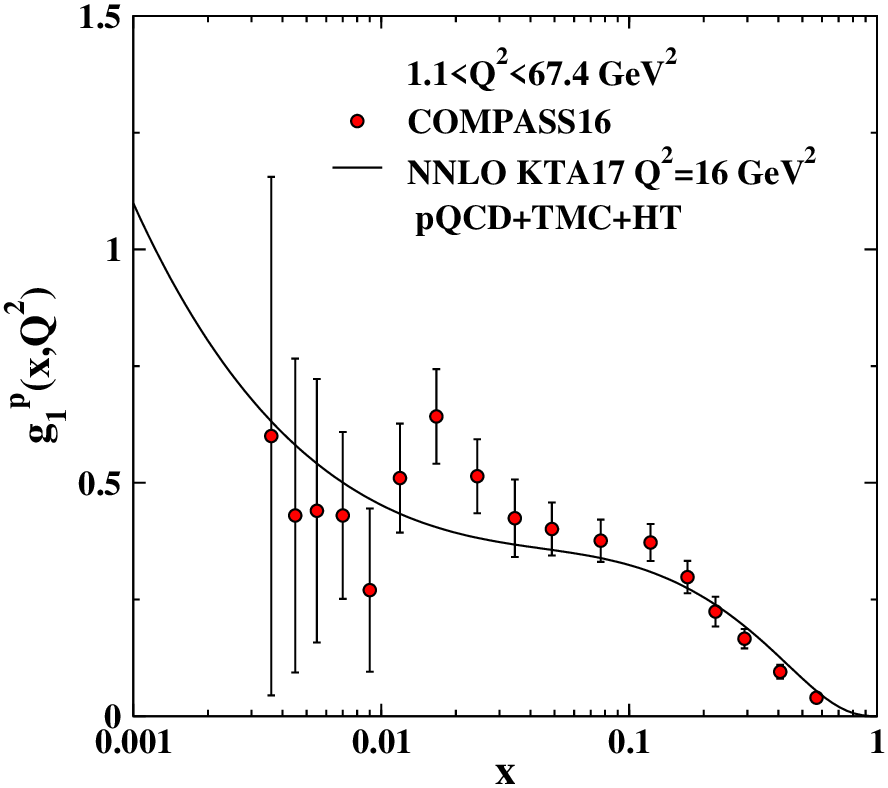}
	\begin{center}
		\caption{ \small (Color online)  KTA17 prediction for the polarized proton structure function $g_1^p$ as a function of $x$ and for mean value of
			$Q^2= 16$ GeV$^2$. Also shown are the most recent data from the
			COMPASS16 collaboration~\cite{Adolph:2015saz}. Note that the values of $Q^2$ for each data point are different. \label{fig:g1compass} }
	\end{center}
\end{figure}

In Figs.~\ref{fig:xg1p}, \ref{fig:xg1n}, and \ref{fig:xg1d},  KTA17 theory predictions for the polarized structure functions of the proton $xg_1^p (x, Q^2)$, neutron $xg_1^n(x, Q^2)$ and deuteron $xg_1^d(x, Q^2)$ are compared with the fixed-target DIS experimental data from E143, E154 and SMC. As we mentioned, KTA17 refers to the pQCD+TMC+HT scenario. The results from KATAO analysis in the NLO approximation~\cite{Khorramian:2010qa} and TKAA16 analysis in the NNLO approximation~\cite{Shahri:2016uzl} also shown. Our curves are presented for some selected values of $Q^2$ = 2, 3, 5, and 10 GeV$^2$ as a function of $x$.
In general, we find good agreement with the experimental data over the entire range of $x$ and Q$^2$, and our results are in accord with other determinations.
In Fig.~\ref{fig:g1compass}, we check the consistency of KTA17 with the newly improved statistical precision data of COMPASS16 in the low-$x$ region.

\begin{figure}[!htb]
	\vspace*{0.5cm}
	\includegraphics[clip,width=0.5\textwidth]{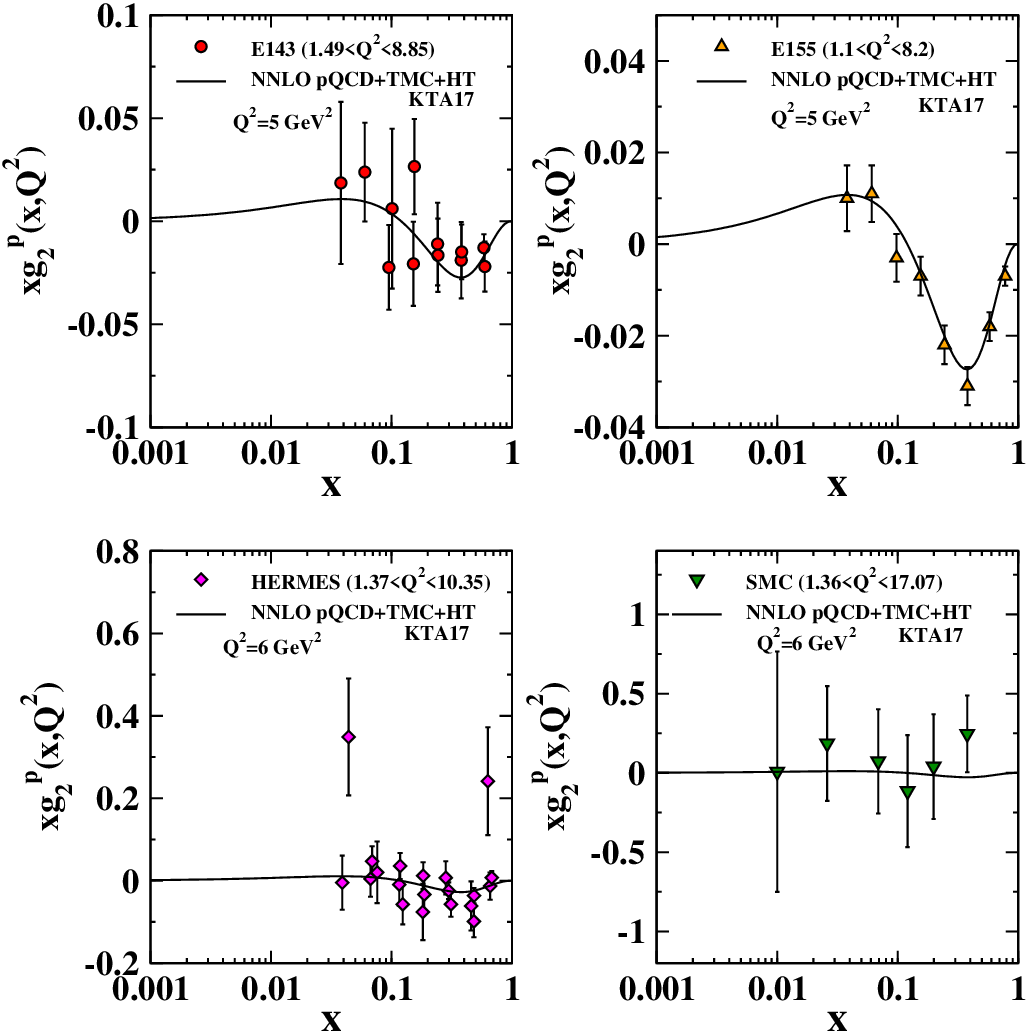}
	\begin{center}
		\caption{ \small (Color online)  KTA17 result for the proton structure functions $xg_2^p(x, Q^2)$ as a function of $x$ and Q$^2$ compared to E143, E155, HERMES and SMC experimental data. \label{fig:xg2p} }
	\end{center}
\end{figure}

\begin{figure}[!htb]
	\vspace*{0.5cm}
	\includegraphics[clip,width=0.5\textwidth]{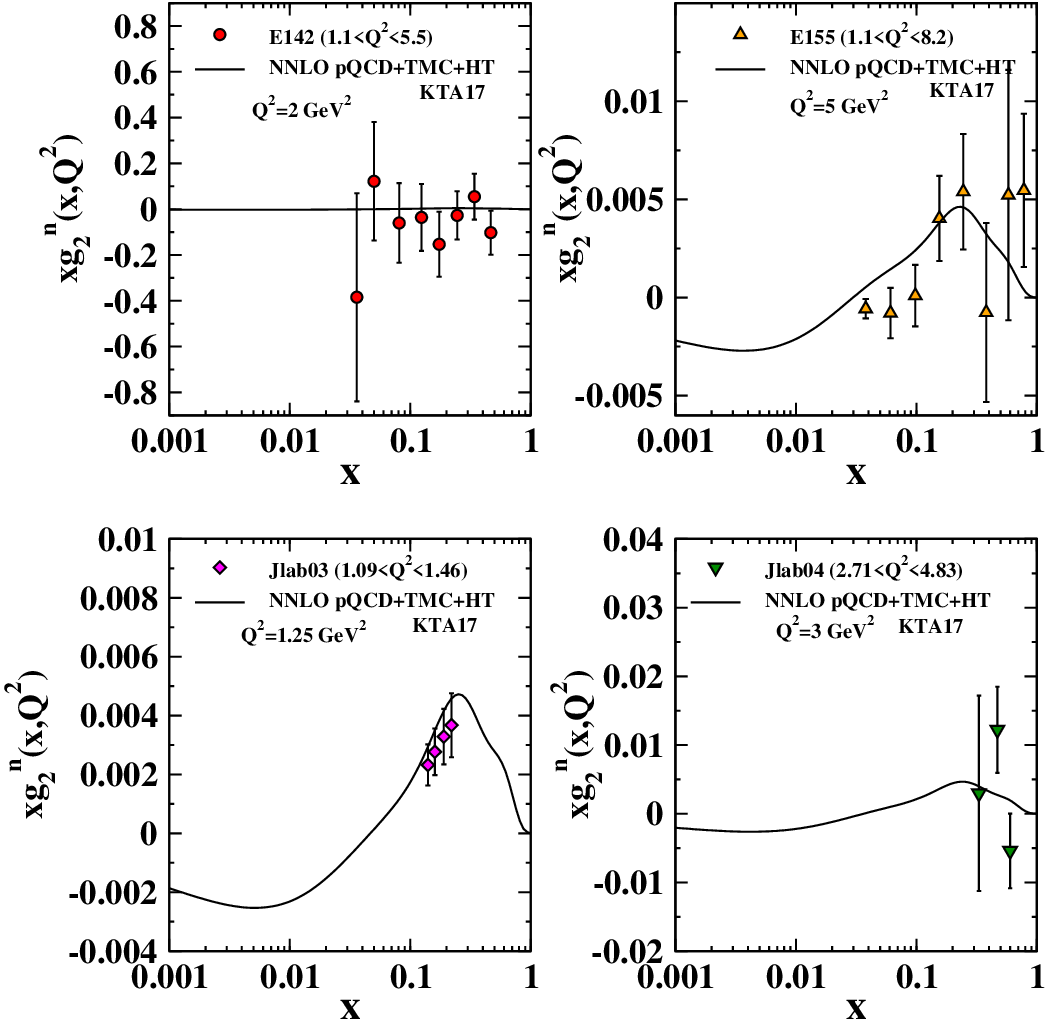}
	\begin{center}
		\caption{ \small (Color online)  KTA17  result for the neutron structure functions $xg_2^n(x, Q^2)$ as a function of $x$ and Q$^2$ compared to E142, E155, JLAB03 and JLAB04 experimental data. \label{fig:xg2n} }
	\end{center}
\end{figure}

\begin{figure}[!htb]
	\vspace*{0.5cm}
	\includegraphics[clip,width=0.5\textwidth]{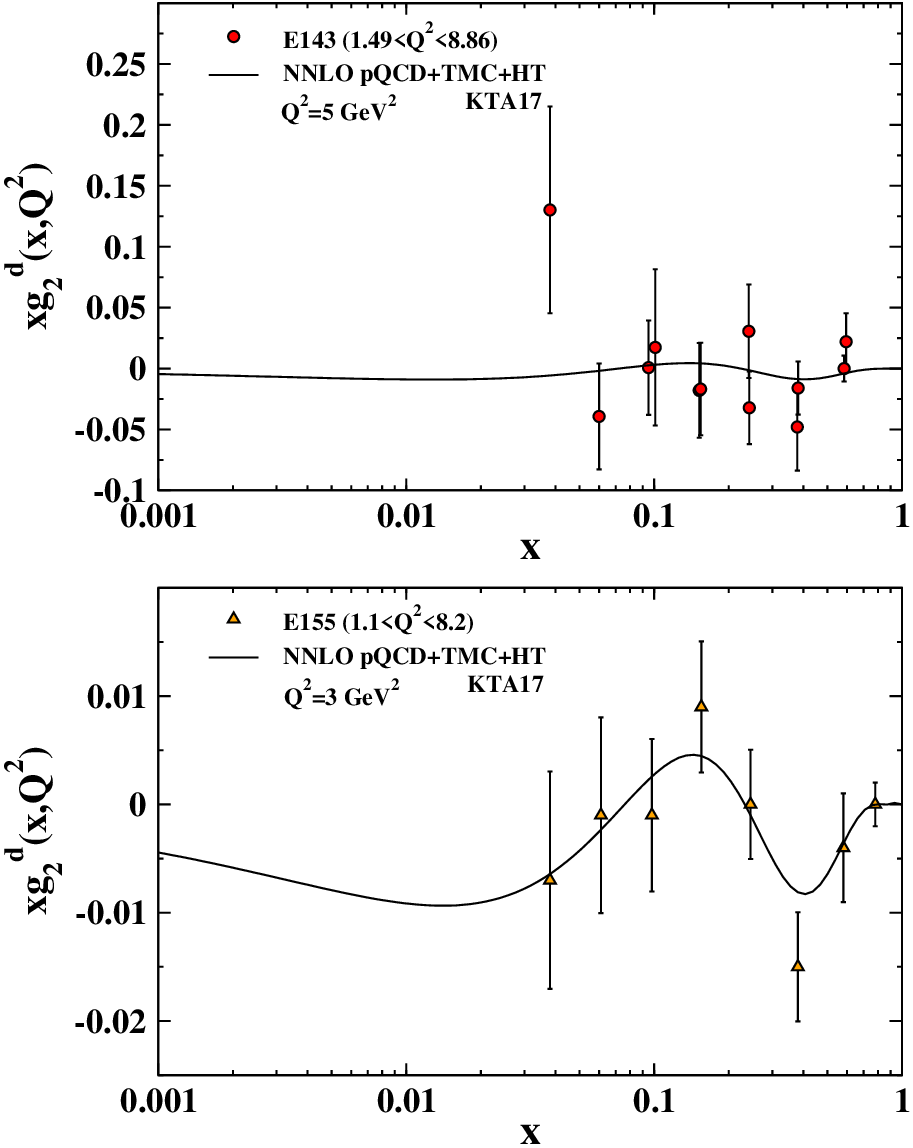}
	\begin{center}
		\caption{ \small (Color online)   KTA17 for the deuteron structure functions $xg_2^d(x, Q^2)$ as a function of $x$ and Q$^2$ compared to E143 and E155 experimental data. \label{fig:xg2d} }
	\end{center}
\end{figure}

Further illustrations of the fit quality are presented in Figs.~\ref{fig:xg2p}, \ref{fig:xg2n} and \ref{fig:xg2d} for the $x g_2^{i = p, n, d}(x, Q^2)$ polarized structure functions obtained from Eq.~\eqref{eq:g1full}.
Generally the $g_2$ data have larger uncertainties compared with the $g_1$ data, reflecting the lack of knowledge in $g_2$ structure function.
At the current level of accuracy,  KTA17  is in agreement with data within their uncertainties, except for the E155 data for $x g_2^d(x, Q^2)$.
A precise quantitative extraction of the $x g_2(x, Q^2)$ requires a large number of data with higher precision.
Our results focus on the general characteristic of the $x g_2(x, Q^2)$.

%
\subsection{ Higher-twist contributions }

\begin{figure}[!htb]
	\vspace*{0.5cm}
	\includegraphics[clip,width=0.45\textwidth]{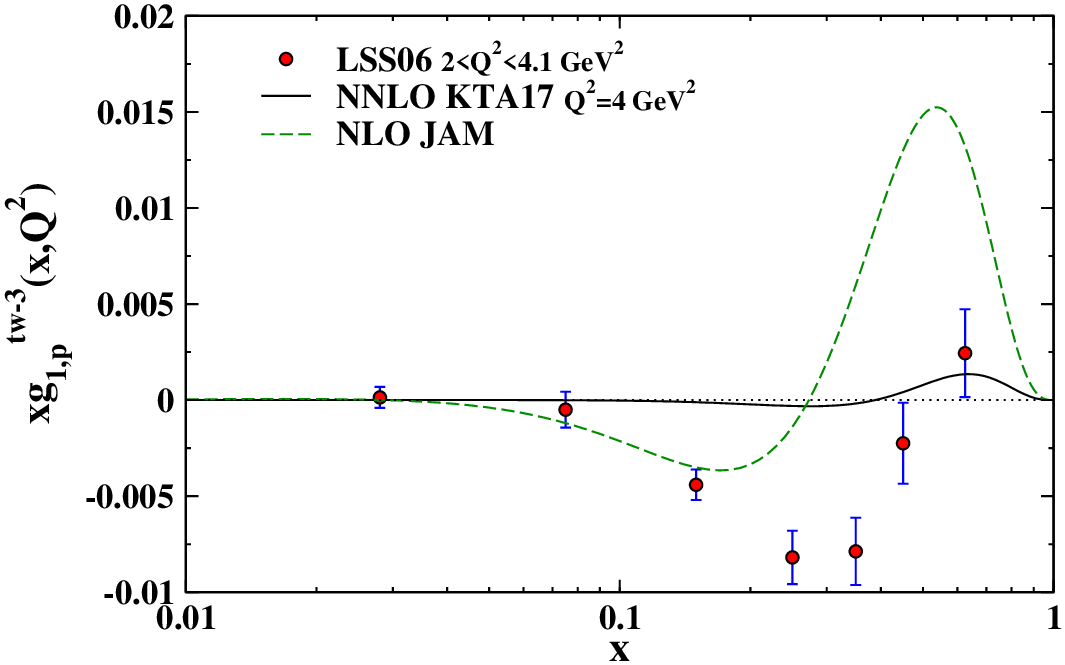}
	\begin{center}
		\caption{ \small (Color online) The twist-3 contribution to $xg_1^p$ at $Q^2$=4 GeV$^2$  as a function of $x$ compared to the results of LSS~\cite{Leader:2006xc} and JAM13~\cite{Jimenez-Delgado:2013boa}. \label{fig:xg1t3Q1} }
	\end{center}
\end{figure}

Figure~\ref{fig:xg1t3Q1} represents our $xg_1^{tw-3}(x,Q^2)$ with those of LSS ~\cite{Leader:2006xc} and JAM13~\cite{Jimenez-Delgado:2013boa}. LSS split the measured $x$ region into seven bins to determine the HT correction to $g_1$. They extracted the HT contribution in a model-independent way, while its scale dependence was ignored.The
JAM group parametrized an analytical form for the twist-3 part of $g_2$ and calculated $g_1^{tw-3}$  by integral relation of Eq.(~\ref{eq:g1HT}) in a global fit at NLO approximation.

\begin{figure}[!htb]
	\vspace*{0.5cm}
	\includegraphics[clip,width=0.45\textwidth]{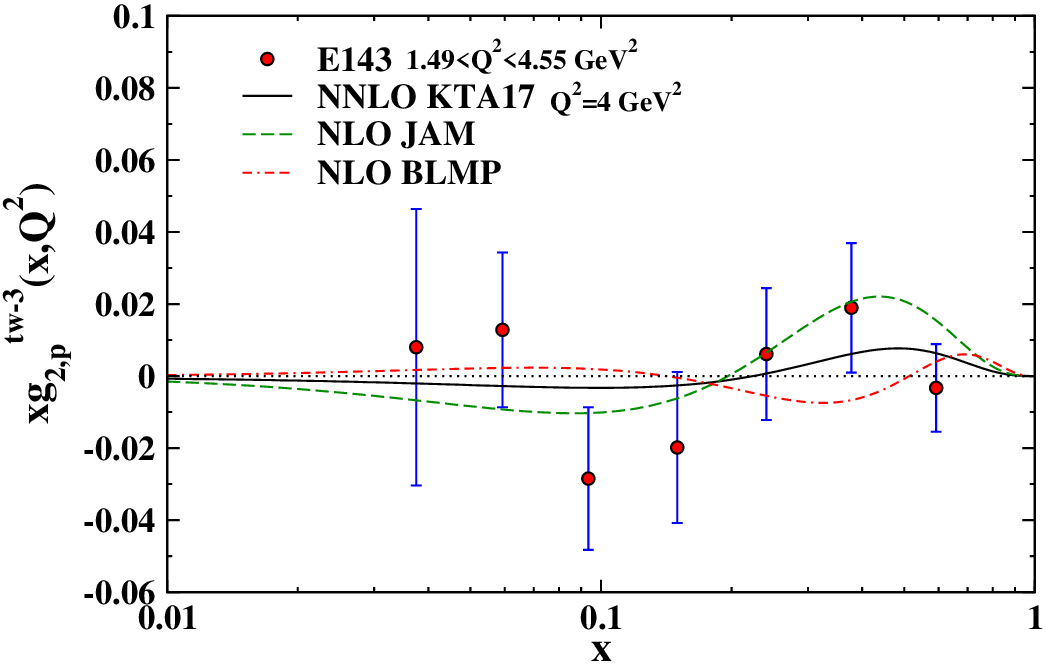}
	\begin{center}
		\caption{ \small (Color online) The twist-3 contribution to $xg_2^p$ at $Q^2$ =4 GeV$^2$  as a function of $x$.  KTA17 (solid curve) is compared with JAM13~\cite{Jimenez-Delgado:2013boa}(dashed), BLMP \cite{Braun:2011aw}
(dashed dotted), and E143 experimental data~\cite{Abe:1998wq}. \label{fig:xg2twQ1} }
	\end{center}
\end{figure}

The twist-3 part of $g_2$ together with those of the JAM13~\cite{Jimenez-Delgado:2013boa} and BLMP~\cite{Braun:2011aw} groups along with E143 experimental data~\cite{Abe:1998wq} are presented in Fig.~\ref{fig:xg2twQ1}.  Keeping terms up to twist 3, E143 Collaboration at SLAC reported the twist 3 contribution to the proton spin structure function $x g_2^p$ with relatively large errors.  However, within experimental precision the $g_2$ data are well described by the twist-2 contribution.
The precision of the current data is not sufficient enough to distinguish model precision.


\begin{figure}[!htb]
	\vspace*{0.5cm}
	\includegraphics[clip,width=0.45\textwidth]{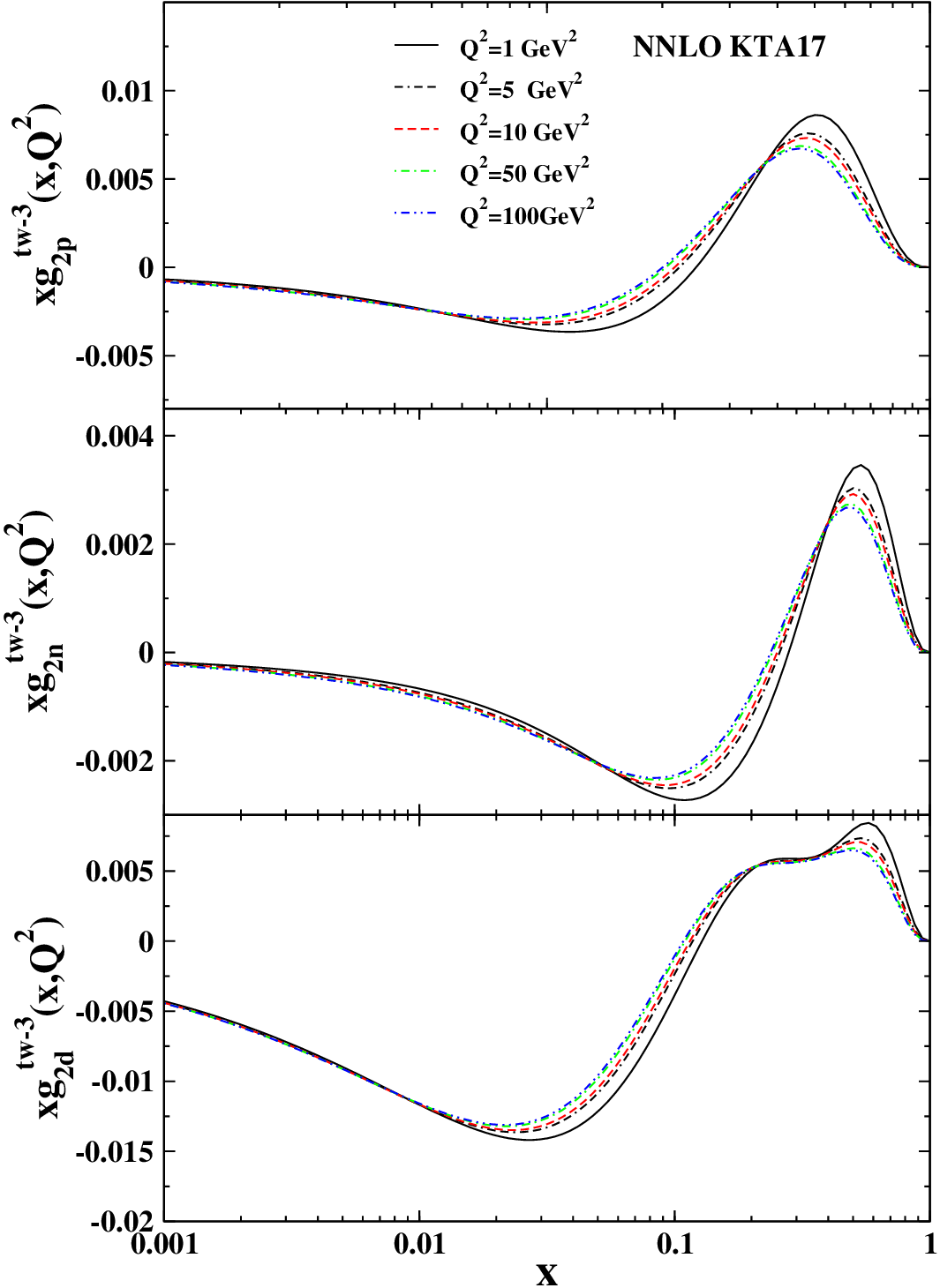}
	\begin{center}
		\caption{\small (Color online) The twist-3 contribution of $xg_2$ for the proton, neutron, and deuteron as a function of $x$ and for different
			values of Q$^2$ according to the KTA17 NNLO analysis. \label{fig:xg2}}
	\end{center}
\end{figure}

\begin{figure}[!htb]
	\vspace*{0.5cm}
	\includegraphics[clip,width=0.45\textwidth]{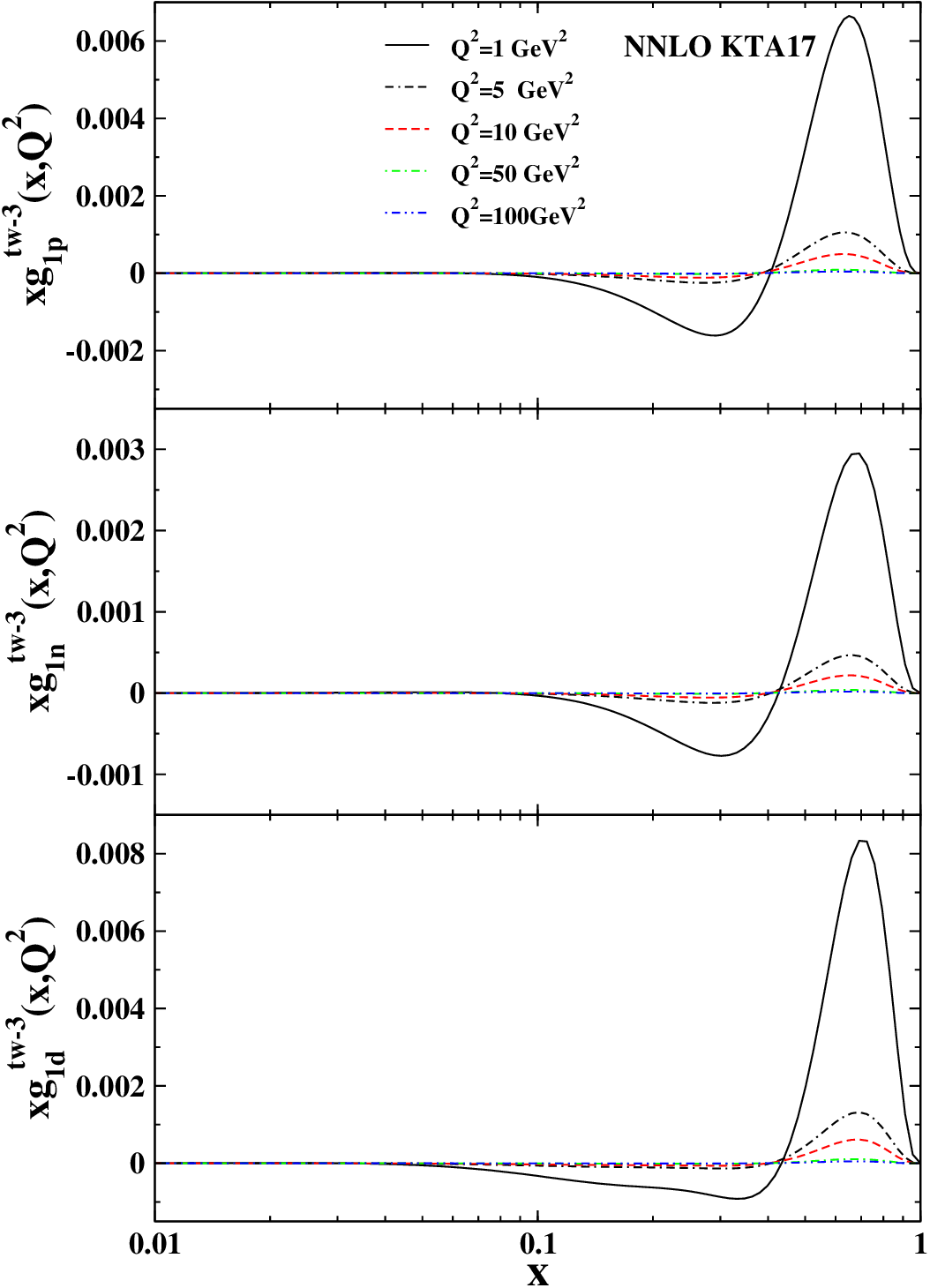}
	\begin{center}
		\caption{\small (Color online) The twist-3 contribution of $xg_1$ for the proton, neutron, and deuteron as a function of $x$ and for different
			values of Q$^2$ according to the KTA17 NNLO analysis. \label{fig:xg1}}
	\end{center}
\end{figure}

As illustrated in Fig.~\ref{fig:xg2}, the twist-3 part of $g_2$ has significant contribution even at large Q$^2$.
In comparison with Fig.~\ref{fig:xg1}, we find that $xg_{\rm 1}^{\tau 3}$ vanishes rapidly at $Q^2 > 5$ GeV$^2$ while $xg_{\rm 2}^{\tau 3}$ remains nonzero even in the limit of Q$^2 \rightarrow \infty$.
%

\begin{figure*}[!htb]
	\vspace*{0.5cm}
	\includegraphics[clip,width=0.9\textwidth]{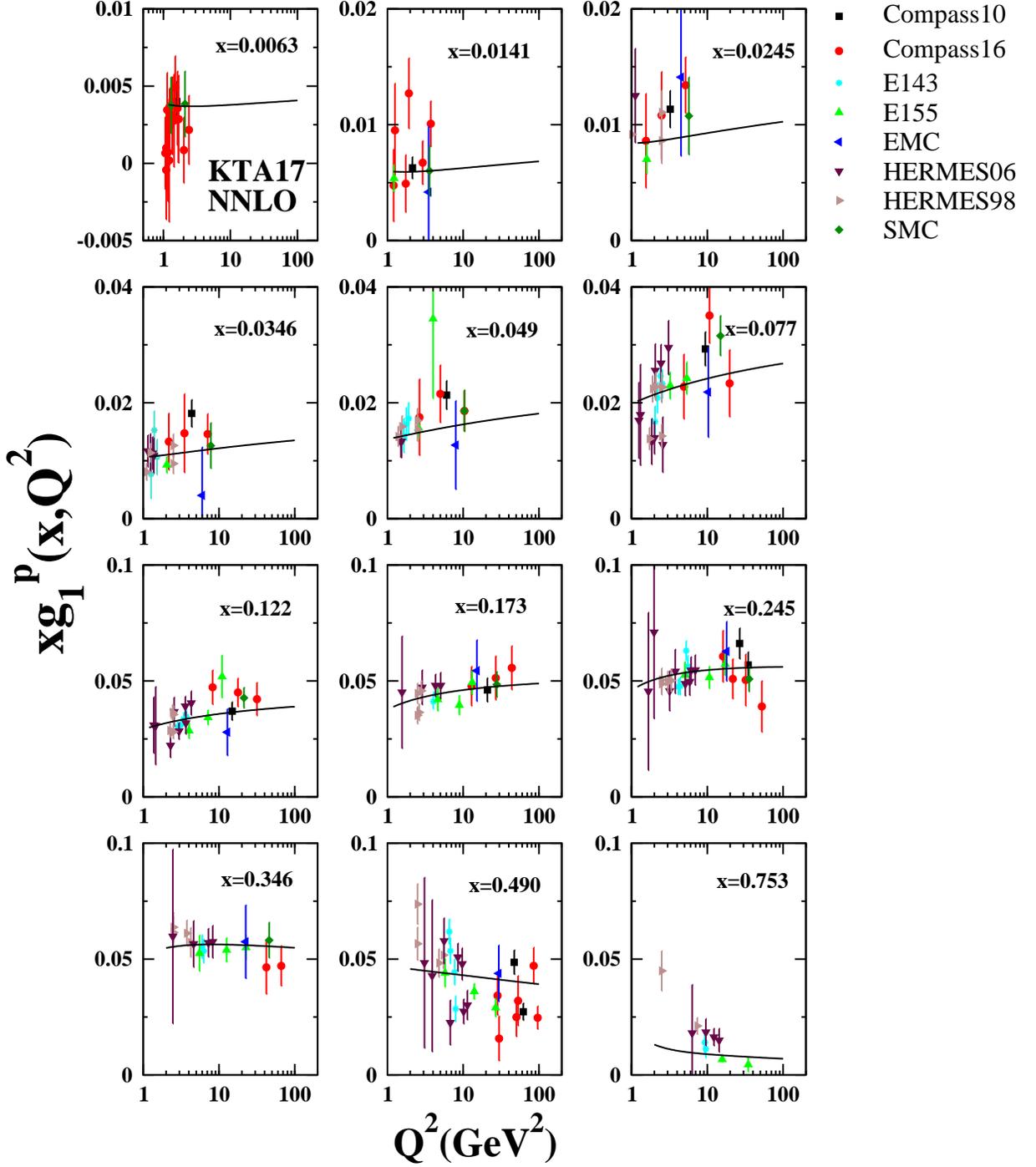}
	\begin{center}
		\caption{\small (Color online)  KTA17  theory predictions for the $xg_1(x, Q^2)$ in comparison to DIS data from Compass10, Compass16, E143, E155, EMC, HERMES06, HERMES98 and SMC experiments.\label{fig:QCDFit-1}}
	\end{center}
\end{figure*}

Finally, KTA17 QCD fit results on $xg_1$ are compared to experimental measurements in Fig.~\ref{fig:QCDFit-1}.
These measurements come from the Compass10, Compass16, E143, E155, EMC, HERMES06, HERMES98 and SMC experiments. The curves are given vs $Q^2$ at several values of $x$ and are compared to the data.  As can be seen, the theory predictions are in good agreement with the data.

%
\section{Sum Rules} \label{Sum-Rules}
%

Sum rules are powerful tools to investigate some fundamental properties of the nucleon structure, like the total momentum fraction carried by partons or the total contribution of parton spin to the spin of the nucleon.
We explore how well the inclusion of TMCs and HT terms into NNLO polarized structure function analysis improves the precision of PPDF determination as well as QCD sum rules.
In the following, the description of almost all important polarized sum rules together with available experimental data are briefly discussed.

%
\subsection{Bjorken sum rule} \label{Bjorken-sum-rule}
%

The nonsinglet spin structure function is defined as
\begin{eqnarray}\label{eq:nonsing}
g_1^{\rm NS}(x, Q^2) &=& g_1^p(x, Q^2) - g_1^n(x,Q^2)\,.
\end{eqnarray}
The polarized Bjorken sum rule expresses the integral over the spin distributions of quarks inside of the nucleon in terms of its axial charge times a coefficient function~\cite{Bjorken:1969mm} as
\begin{eqnarray}\label{eq:Bjorken-SR}
\Gamma_1^{\rm NS}(Q^2)&=&\Gamma_1^p(Q^2) - \Gamma_1^n(Q^2)   \nonumber \\
&=& \int_0^1[g_{1}^{p}(x, Q^2) - g_1^{{n}}(x, Q^2)]dx        \nonumber \\
&=& \frac{1}{6} ~ |g_A|~ C_{Bj}[\alpha_s(Q^2)] + \text{HT corrections}\,.\nonumber \\
\end{eqnarray}
Here, $g_A$ is the nucleon axial charge as measured in neutron $\beta$ decay.
The coefficient function $C_{Bj}[\alpha_s(Q^2)]$ is calculated in four-loop pQCD corrections in the massless \cite{Baikov:2010je} and very recently massive cases \cite{Blumlein:2016xcy}.
Bjorken sum rule potentially provides a very precise handle on the strong coupling constant. The value of $\alpha_s$ can be extracted via $C_{Bj}[\alpha_s(Q^2)]$ expression from experimental data. $\alpha_s$ is also available form accurate methods, such as the $\tau$ lepton and the $Z$ boson into hadrons width decay. Comparison of these values offers an important test of QCD consistency.
As previously reported in Ref.~\cite{Altarelli:1998nb}, determination of $\alpha_s$ from the Bjorken sum rule suffers from small-$x$ extrapolation ambiguities.

Our results for the Bjorken sum rule are compared with experimental measurements \text{E143}~\cite{Abe:1998wq}, \text{SMC}~\cite{SMCpg2}, \text{HERMES06}~\cite{HERMpd} and \text{COMPASS16}~\cite{Adolph:2015saz} in Table~\ref{tab:Bjorken}.
%
%
\begin{table*}[!htb]
	\caption{\label{tab:Bjorken} Comparison of the result of the Bjorken sum rule for $\Gamma_1^{NS}$ with world data from
		\text{E143}~\cite{Abe:1998wq}, \text{SMC}~\cite{SMCpg2}, \text{HERMES06}~\cite{HERMpd} and \text{COMPASS16}~\cite{Adolph:2015saz}.
		Only HERMES06~\cite{HERMpd} results are not extrapolated in full $x$ range (measured in region $0.021 \leq x \leq 0.9$). }
	\begin{ruledtabular}
	     \begin{tabular}{lcccccc}
& \textbf{E143}~\cite{Abe:1998wq}   & \textbf{SMC}~\cite{SMCpg2}  & \textbf{HERMES06}~\cite{HERMpd}  &  \textbf{COMPASS16}~\cite{Adolph:2015saz} & \textbf{KTA17} \\ 
& $Q^2=5$ GeV$^2$& $Q^2=5$ GeV$^2$& $Q^2=5$ GeV$^2$& $Q^2=3$ GeV$^2$&$Q^2=5$ GeV$^2$       \\     \hline    \hline\tabularnewline
$\Gamma^{\rm NS}_1$   & $0.164 \pm 0.021$ & $ 0.181\pm 0.035 $  & $0.148 \pm 0.017$ & $0.181\pm 0.008$& $0.173\pm0.003$     \\
		\end{tabular}
	\end{ruledtabular}
\end{table*}

%
\subsection{Proton helicity sum rule}
%

The extrapolation of the proton spin among its constituents is a compelling question still driving the field of nuclear physics \cite{Leader:2016sli}.
In order to get an accurate picture of the quark and gluon helicity density a precise extraction of PPDFs entering the proton's momentum sum rule is required.
In a general approach, the spin of the nucleon can be carried by its constituents as
\begin{equation}\label{eq:spinsumrule}
\frac{1}{2} = \frac{1}{2} \Delta \Sigma(Q^2) + \Delta {\mathrm G}(Q^2) + {\mathrm L}(Q^2).
\end{equation}
Here, $\Delta {\rm G(Q^2)}=\int_{0}^{1}dx~\Delta g(x,Q^2)$ has the interpretation of the gluon spin contribution, and $\Delta \Sigma(Q^2)=\sum_{i}\int_{0}^{1}dx~(\Delta q(x,Q^2)+\Delta \bar{q}(x,Q^2))$ denotes the flavor singlet spin contribution.
${\mathrm L}(Q^2)$ is the total contribution from the quark and gluon orbital angular momentum.
Finding a way to measure them is a real challenge beyond the scope of this paper.  Each individual term in  Eq. (\ref{eq:spinsumrule})  is a function of $Q^2$n but the sum is not.
The values of the singlet-quark and gluon first moment at the scale of Q$^2$=10 GeV$^2$ are listed in Table~\ref{tab:firstmomentQ10}. Results are compared to those from the \text{NNPDFpol1.0}~\cite{,Ball:2013lla}, \text{NNPDFpol1.1}~\cite{Nocera:2012hx} and \text{DSSV08}~\cite{deFlorian:2008mr} at both the truncated and full $x$ regions.
In Table~\ref{tab:firstmomentQ4}, KTA17 results are presented and compared at Q$^2$=4 GeV$^2$ with the \text{DSSV08}~\cite{deFlorian:2008mr}, \text{BB10}~\cite{Blumlein:2010rn}, \text{LSS10}~\cite{Leader:2010rb} and \text{NNPDFpol1.0}~\cite{,Ball:2013lla} results.

Coming now to a comparison of results, we see that for the $\Delta \Sigma$, KTA17 results are consistent within uncertainties with that of other groups.
This is mainly because the first moment of polarized densities is fixed by semileptonic decays.
Turning to the gluon, very different values are reported. The large uncertainty  prevents reaching a firm conclusion about the full first moment of the gluon.

%
\begin{table*}[!htb]
	\caption{\label{tab:firstmomentQ10} Results for the full and truncated first moments of the polarized singlet-quark $\Delta \Sigma(Q^2)=\sum_{i}\int_0^1 dx [\Delta q_i(x) + \Delta \bar q_i (x)]$ and gluon distributions at the scale Q$^2$=10 GeV$^2$ in the $\overline{{\rm MS}}$ scheme. Also shown are the recent polarized global analyses of \text{NNPDFpol1.0}~\cite{,Ball:2013lla}, \text{NNPDFpol1.1}~\cite{Nocera:2012hx} and \text{DSSV08}~\cite{deFlorian:2008mr}. }
	\begin{ruledtabular}
		\begin{tabular}{lccccc}
			& \textbf{DSSV08}~\cite{deFlorian:2008mr}  & \textbf{NNPDFpol1.0}~\cite{,Ball:2013lla} & \textbf{NNPDFpol1.1}~\cite{Nocera:2012hx}&  \textbf{KTA17}     \tabularnewline
			Full $x$ region $[0,1]$ &  &  & &  \\
			\hline \hline \tabularnewline
			$\Delta \Sigma {\rm(Q^2)}$&$0.242$  &$+0.16\pm 0.30$ &$+0.18\pm 0.21$       &$0.210\pm0.045$  \\
			$\Delta {\rm G(Q^2)}$   & $-0.084$ & $-0.95\pm 3.87$ & $0.03\pm 3.24$     &$0.138\pm0.058$  \\ \hline  \hline  \tabularnewline
Truncated $x$ region [$10^{-3},1$] &  &  & &  \\     \hline \hline \tabularnewline
	$\Delta \Sigma {\rm(Q^2)}$  & $0.366\pm0.017$& $+0.23\pm0.15$&$+0.25\pm0.10$    &$0.234\pm0.044$ \\
	$\Delta {\rm G(Q^2)}$   &$0.013\pm0.182$  & $-0.06\pm 1.12$ & $0.49\pm 0.75$  &$0.138\pm0.058$
		\end{tabular}
	\end{ruledtabular}
\end{table*}
%
%
\begin{table*}[!htb]
	\caption{\label{tab:firstmomentQ4} Same as Table~\ref{tab:firstmomentQ10}, but only for the full first moments of the polarized singlet-quark and gluon distributions at the scale Q$^2$ =4 GeV$^2$ in the $\overline{{\rm MS}}$ scheme. Those of \text{DSSV08}~\cite{deFlorian:2008mr}, \text{BB10}~\cite{Blumlein:2010rn}, \text{LSS10}~\cite{Leader:2010rb} and \text{NNPDFpol1.0}~\cite{,Ball:2013lla} are presented for comparison. }
	\begin{ruledtabular}
		\begin{tabular}{ccccccc}
			& \textbf{DSSV08}~\cite{deFlorian:2008mr}  & \textbf{BB10}~\cite{Blumlein:2010rn} & \textbf{LSS10}~\cite{Leader:2010rb} & \textbf{NNPDFpol1.0}~\cite{,Ball:2013lla} &  \textbf{KTA17}     \tabularnewline
			\hline\hline
			$\Delta \Sigma {\rm(Q^2)}$  & $0.245$ & $0.193 \pm  0.075$ & $0.207 \pm 0.034$ & $0.18 \pm 0.20$ & $0.232\pm0.044$  \\
			$\Delta {\rm G(Q^2)}$   & $-0.096$ & $0.462 \pm 0.430$ & $0.316 \pm 0.190$ & $-0.9 \pm 4.2$ & $0.128\pm0.053$   \\
		\end{tabular}
	\end{ruledtabular}
\end{table*}

Let us finally discus the proton spin sum rule based on the extracted values presented in Table~\ref{tab:firstmomentQ4}. The total orbital angular momentum to the total spin of the proton is
\begin{equation}
{\mathrm L}(Q^2=4~\rm GeV^2) = 
0.256 \pm 0.069 \,.
\end{equation}
The gluon uncertainty is clearly dominant. Due to large uncertainty originating mainly from the gluons, we cannot yet come to a definite conclusion about the contribution of the total orbital angular momentum to the spin of the proton.
Improving the current level of experimental accuracy is required for the precise determination of each individual contribution.

%
\subsection{twist 3 reduced matrix element $d_2$}

Under the OPE, one can study the effect of quark-gluon correlations via the moments of $g_1$ and $g_2$
\begin{eqnarray}
d_2(Q^2) &=& 3 \int_0^1 x^2 \bar{g_2}(x,Q^2)~dx       \nonumber   \\
&=& \int_0^1 x^2 [3 g_2(x, Q^2) + 2g_1(x, Q^2)]~ dx,
\end{eqnarray}
as follows from the relation $\bar g_2=g_2-g_{2} ^{WW}$.
Thus, the twist-3 reduced matrix element of spin-dependent operators in the nucleon measures the deviation of $g_2$ from $g_{2} ^{\tau 2}$ [See Eq.~\eqref{WW}].
The function of $d_2(Q^2)$  is especially sensitive to the large-$x$ behavior of $\bar{g_2}$ (due to the $x^2$ weighting factor).
Extraction of $d_2$ is particularly interesting as it will provide insight into the size of the multiparton correlation terms.

Our results together with the other theoretical and experimental values are presented in Table~\ref{tab:twist3}.
This notably nonzero value for $d_2$ implies the significance of considering higher-twist terms in QCD analyses.
The most reliable determination of the the higher-twist moments $d_2$ was performed in JAM15~\cite{Sato:2016tuz}. Since they are the only group that implemented TMCs for the $\tau3$ part.

In the near future, the expected data from 12 GeV Jefferson Lab experiments~\cite{JLAB-12}  may enable the $d_2$ moments to be determined more precisely in the DIS region at higher $Q^2$ values.
QCD analysis of this new generation of bounded uncertainty data requires including TMCs in all higher-twist terms.

%
%
%
\begin{table*}[!htb]
 	\caption{\label{tab:twist3} $d_2$ moments of the proton, neutron and deuteron polarized structure functions from the \text{SLAC E155x}~\cite{Kuhn:2008sy},
\text{E01-012}~\cite{Solvignon:2013yun}, \text{E06-014}~\cite{Flay:2016wie}, \text{Lattice QCD}~\cite{Gockeler:2005vw}, \text{CM bag model}~\cite{Song:1996ea}, \text{JAM15}~\cite{Sato:2016tuz}, and \text{JAM13}~\cite{Jimenez-Delgado:2013boa} compared with  KTA17 results.
 	}
\begin{ruledtabular}
 		\begin{tabular}{lccccc}
& Ref. &  \textbf{$Q^2$ [GeV$^{2}$]} &  \textbf{$10^2d^{p}_2$}  & \textbf{$10^5d^{n}_2$}  & \textbf{$10^3d^d_2$} \tabularnewline  \hline  \hline  \tabularnewline
KTA17      &&$5$&$0.66\pm0.01$         & $193.81\pm6.42$        & $6.97\pm0.11$       \\
\text{E06-014} & \cite{Flay:2016wie}& 3.21 & &$-421.0 \pm 79.0 \pm 82.0 \pm 8.0$ & -\\
\text{E06-014} & \cite{Flay:2016wie}& 4.32 & &$-35.0 \pm 83.0 \pm 69.0 \pm 7.0$  & -\\
\text{E01-012} & \cite{Solvignon:2013yun} &3 & - & $-117 \pm 88 \pm 138$ &  -  \\
\text{E155x}   & \cite{E155pdg2} & $5$ & $0.32\pm 0.17$ & $790\pm 480$ &     -                         \\
\text{E143}    & \cite{Abe:1998wq}& $5$                 & $0.58\pm 0.50$ & $500\pm 2100$ & $5.1\pm 9.2$ \\
\text{Lattice QCD}  &\cite{Gockeler:2005vw} & 5 &  0.4(5) & -100(-300) & - \\
\text{CM bag model} &\cite{Song:1996ea} & $5$  & $1.74$          & $-253$       & $6.79 $         \\
\text{JAM15}        &\cite{Sato:2016tuz} & $1$  & $0.5\pm 0.2$          & $-100 \pm 100$       & -        \\
\text{JAM13}        &\cite{Jimenez-Delgado:2013boa} & $5$  & $1.1 \pm 0.2$          & $200 \pm 300$       &  -         \\

 		\end{tabular}
 \end{ruledtabular}
 \end{table*}
%

%
\subsection{Burkhardt-Cottingham (BC) sum rule} \label{BC-sum-rule}
%

The first moment of $g_2$ is predicted to yield zero by Burkhardt and Cottingham (BC) from virtual Compton scattering dispersion relations in all Q$^2$~\cite{Burkhardt:1970ti}
\begin{equation}
\Gamma_2 = \int_0^1 dx \, g_2(x, Q^2) = 0~.
\end{equation}
It appears to be a trivial consequence of the WW relation for $g_{2}^{\tau2}$.
The BC sum rule is also satisfied for the target mass corrected structure functions.
Therefore a violation of the BC sum rule would imply the presence of HT contributions~\cite{hermes2012g2}.
Our $\Gamma_2$ results together with data from the \text{E143}~\cite{Abe:1998wq}, \text{E155}~\cite{E155pdg2}, \text{HERMES2012}~\cite{hermes2012g2}, \text{RSS}~\cite{Slifer:2008xu}, and \text{E01012}~\cite{Solvignon:2013yun} groups for the proton, deuteron and neutron are presented in Table~\ref{tab:BC}. Any conclusion depends on the low-$x$ behavior of $g_2$ which has not yet been precisely measured.

%
%
\begin{table*}[!htb]
	\caption{ \label{tab:BC} Comparison of the result of the BC sum rule for $\Gamma_2^p$, $\Gamma_2^d$ and $\Gamma_2^n$ with world data from \text{E143}~\cite{Abe:1998wq}, \text{E155}~\cite{E155pdg2}, \text{HERMES2012}~\cite{hermes2012g2}, \text{RSS}~\cite{Slifer:2008xu}, and \text{E01012}~\cite{Solvignon:2013yun}. }
	\begin{ruledtabular}
		\begin{tabular}{lccccccc}
			
			& \textbf{E143}~\cite{Abe:1998wq}  & \textbf{E155}~\cite{E155pdg2}  & \textbf{HERMES2012}~\cite{hermes2012g2}& \textbf{RSS}~\cite{Slifer:2008xu} & \textbf{E01012}~\cite{Solvignon:2013yun} & \textbf{KTA17}   \\
			& $0.03 \le x \le 1 $  & $0.02 \le x \le 0.8 $   & $0.023 \le x \le 0.9$  &$0.316 < x < 0.823 $&$0 \le x \le 1 $& $0.03 \le x \le 1 $\\
			& $Q^2=5$ GeV$^2$& $Q^2=5$ GeV$^2$& $Q^2=5$ GeV$^2$& $Q^2=1.28$ GeV$^2$&$Q^2=3$ GeV$^2$&$Q^2=5$ GeV$^2$   \\
			\hline  \hline
			$\Gamma_2^p$  & $-0.014 \pm 0.028$ &$-0.044 \pm 0.008$ & $0.006 \pm 0.029$ & $-0.0006 \pm 0.0022$&...&  $ -0.0171 \pm 0.0004$   \\
			$\Gamma_2^d$  & $-0.034 \pm 0.082$ &$-0.008 \pm 0.012$ &-  &$-0.0090 \pm 0.0026$&...& $ -0.0051 \pm 0.0008$    \\
			$\Gamma_2^n$ &-&-&-&$-0.0092 \pm 0.0035$&$0.00015 \pm 0.00113$& $ ~0.0080 \pm 0.0013$  \\
		\end{tabular}
	\end{ruledtabular}
\end{table*}

%
\subsection{Efremov-Leader-Teryaev sum rule}

The Efremov-Leader-Teryaev (ELT) sum rule~\cite{Efremov:1996hd} integrates the valence part of $g_1$ and $g_2$ over $x$.
Considering that the sea quarks are the same in protons and neutrons, the ELT sum rule can be derived similar to the Bjorken sum rule as
\begin{eqnarray}
&&\int_0^1 dx ~ x[g_1^V(x) + 2 g_2^V (x)]=  \nonumber \\
&&\int_0^1 dx ~ x[g_1^p(x) - g_1^n(x) + 2(g_2^p(x) - g_2^n(x))]=0.
\end{eqnarray}
This sum rule receives quark mass corrections and is only valid in the case of massless quarks \cite{Blumlein:1996vs}.
It is preserved under the presence of target mass corrections~\cite{Blumlein:1998nv}.
Combining the data of E143~\cite{Abe:1998wq} and E155~\cite{E155pdg2} leads to $-0.011 \pm 0.008$ at Q$^2$=5 GeV$^2$. We extracted the value of $0.0063\pm 0.0003$ at the same Q$^2$.

%
%
\section{ Polarized PDFs in the high-precision era of collider physics } \label{high-precision-era}

Several determinations of polarized PDFs of the proton are presently available up to NLO ~\cite{deFlorian:2009vb,Hirai:2008aj,Blumlein:2010rn,Leader:2010rb,Nocera:2014gqa,Jimenez-Delgado:2013boa,Jimenez-Delgado:2014xza,Sato:2016tuz,Arbabifar:2013tma,Monfared:2014nta} and also for the NNLO approximation~\cite{Shahri:2016uzl}.
They mostly differ in the included polarized data sets, the procedure applied to determine PPDFs from these data sets and the method used to extract corresponding uncertainties. Most of the analyses focused on the Lagrange multiplier or the Hessian approaches to estimate the uncertainty, while the NNPDF collaboration has developed a Monte Carlo methodology to control uncertainties.
Available analyses use experimental information from neutral-current DIS and SIDIS to constrain the total quark combinations and individual quark and antiquark flavors, respectively.
The gluon distribution would be constrained rather weakly by both DIS and SIDIS data,  because of the small $Q^2$ range covered.

In addition to the DIS and SIDIS fixed-target data, a remarkable amount of data from longitudinally polarized proton-proton collisions at the RHIC has become available recently~\cite{Aschenauer:2016our,Akiba:2015jwa}. The RHIC data can be expected to further constrain the gluon helicity distribution especially at the small momentum fractions, down to  $x\sim 0.01$~\cite{Aschenauer:2013woa,Aschenauer:2015eha,Ball:2013tyh,Adare:2015ozj}.
The double-helicity asymmetries for jet and $\pi^0$ production are directly sensitive to the gluon helicity distribution over a small range of $x$, because of the dominance of gluon-gluon and quark-gluon initiated subprocesses in the kinematic range accessed by PHENIX at the RHIC~\cite{Adare:2014wht}.
In recent helicity PDF fits~\cite{deFlorian:2008mr,deFlorian:2009vb,Nocera:2014gqa,deFlorian:2014yva}, the  RHIC measurements on the double-longitudinal spin
asymmetry in the production of hadrons~\cite{Adare:2008aa,Adare:2008qb} and inclusive jet production in $pp$ collisions~\cite{Abelev:2007vt}, as well as single-longitudinal spin asymmetry measurements in the production of $W^{\pm}$ bosons~\cite{Aggarwal:2010vc,Adare:2010xa,Nocera:2016xhb}, have already been used.
These data can increase sensitivity to the sign information of gluon density in present and future pQCD helicity PDF fits.
In addition to the mentioned data, inclusion of the Hall-A and CLASS measurements at JLAB leads to a reduction
in the PDF errors for the polarized valence and sea quarks densities as well as the gluon polarization uncertainty at $x\geqslant 0.1$~\cite{Sato:2016tuz}.

The COMPASS Collaboration at CERN performed new measurements of the longitudinal double-spin asymmetry and the longitudinal spin structure function of the proton~\cite{Adolph:2015saz} as well as deuteron~\cite{Adolph:2016myg}.
COMPASS measurements provide the lowest accessible values for $x$ and the largest $Q^2$ values for any given $x$. Consequently, it leads to a better determination of sea quarks and gluon helicity distribution including the corresponding uncertainties.
These data improve the statistical precision of $g^p_1(x)$ by about a factor of 2 in the region $x\leqslant 0.02$.

Despite the discussed achievements, the QCD analysis of polarized data suffers from both limited kinematic coverage and insufficient precision of the available inclusive data. Consequently our understanding of the nucleon spin structure is still far from complete. The most up-to-date 200 GeV data from the COMPASS16 experiment do not change much the general trend of the polarized PDFs but a reduction of the uncertainties on almost all parton species was observed.

Finally, it should be stressed that a future polarized electron-ion collider (EIC) would allow for a major breakthrough toward the understanding of the proton spin. The EIC is expected to open up the kinematic domain to significantly lower values of $x$ ($x \approx 10^{-4}$) in center-of-mass energy to $\sim 104 $ GeV$^2$, reducing significantly the uncertainty on the contributions from the unmeasured small-$x$ region. The EIC will likely be the only facility to study the spin structure of the proton with the highest precision~\cite{Chu:2016wsq,Accardi:2012qut,Nocera:2016xhb,Chudakov:2016ytj,Montgomery:2016uws,Cosyn:2016oiq,Cosyn:2014zfa,Guzey:2014jva,Hoecker:2016vvy}.

%
%
\section{Summary and conclusions} \label{Summary}

The main goal of the present KTA17  analysis is to determine the nucleon spin structure functions $g_1(x,Q^2)$ and $g_2(x,Q^2)$ and their moments which are essential in testing QCD sum rules.
We have enriched our recent NNLO formalism \cite{Shahri:2016uzl} by TMCs and HT terms and extended it to include more experimental observables. These corrections play a significant role in the large-$x$ region at low $Q^2$.
We achieved an excellent description of the fitted data and provided  unified and consistent PPDFs. Our helicity distributions have compared reasonably well with other extractions, within the known very large uncertainties arising from the lack of constraining data.
We also studied the TMCs and HT effects on several sum rules at the NNLO approximation, since they are relevant in the region of low $Q^2$.
The Bjorken sum rule is related to polarized  $g_1$ structure functions.
We also present our results for the reduced matrix element $d_2$ in the NNLO approximation.
More accurate data are required to scrutinize the BC and ELT sum rules.

The future polarized EIC will make a huge impact on our knowledge of spin physics. The decreased uncertainties would absolutely solve the question of how spin and the orbital angular momentum of partons contribute to the overall proton spin.
Concluding, in the light of upcoming development in experimental projects, phenomenological efforts to increase our knowledge of the nucleon structure functions and their moments are particularly important.

%
%
\section*{Acknowledgments}

We would like to thank Elliott Leader and Emanuele Nocera for reading the manuscript and helpful discussions. We thank Alberto Accardi for detailed discussion on the evolution of higher-twist terms and Fabienne Kunne for detailed comments on COMPASS16 polarized DIS data.
We are also thankful for School of Particles and Accelerators, Institute for Research in Fundamental Sciences for financially support this project. Hamzeh Khanpour acknowledges the University of Science and Technology of Mazandaran for financial support provided for this research and is grateful for the hospitality of the Theory Division at CERN where this work has been completed.
S. Taheri Monfared gratefully acknowledges partial support of this research provided by the Islamic Azad University Central Tehran Branch.

\appendix

%
%
\section{FORTRAN package of KTA17 NNLO polarized PDFs}\label{AppendixA}

A FORTRAN package containing  KTA17 NNLO spin-dependent PDFs as well as the polarized structure functions $x\,g_1^{i = p, n, d}(x, Q^2)$ for the proton, neutron and deuteron can be obtained via Email from the authors upon request. This package also includes an example program to illustrate the use of the routines.

%
%
\section{NNLO splitting function}\label{AppendixB}

\begin{widetext}
	
In this section, for completeness, we present the NNLO Mellin-N space splitting function used for the evolution of longitudinally polarized parton densities based on our analysis. Their $x$-space forms are available in  Ref.~\cite{Moch:2014sna}. FORTRAN files of our analytical results can be obtained from the authors upon request.

These function can be written in terms of the harmonic sums as~\cite{Vermaseren:1998uu,Blumlein:1998if},
\begin{eqnarray}
s_1 &=& \gamma _E + \psi (n+1)\,, \nonumber\\
s_2 &=& \zeta (2) - \psi  '(n+1)\,, \nonumber\\
s_3 &=& \zeta (3) + 0.5 \, \psi''(n+1)\,, \nonumber\\
s_4 &=& \zeta (4) - 1/6 \,  \psi'''(n+1), \nonumber
\end{eqnarray}
where $\gamma _E = 0.577216$ is the Euler constant, $\psi(n) = d\ln\Gamma(n)/dn$ is the digamma function and $\zeta (2) = \pi^2/6$, $\zeta (3) = 1.20206$ and $\zeta(4) = 1.08232$.

The analytical expressions for the polarized NNLO quark-quark splitting function are given by
\begin{eqnarray}
&&\Delta p_{qq}^{(2)}=1295.47+\frac{928}{27 n^5}-\frac{640}{3 n^4}+\frac{798.4}{n^3}-\frac{1465.2}{n^2}+\frac{1860.2}{n}-\frac{3505}{1+n}+\frac{297}{2+n}-\frac{433.2}{3+n}+\nonumber\\
&&1174.898\left(\frac{1}{n}-s_1\right)-\frac{714.1 s_1}{n}+684\left(\frac{s_1}{n^2}+\frac{-\zeta (2)+s_2}{n}\right)+\nonumber\\
&&f \left(-173.933+\frac{512}{27 n^4}-\frac{2144}{27 n^3}+\frac{172.69}{n^2}-\frac{216.62}{n}+\frac{6.816}{(1+n)^4}+\frac{406.5}{1+n}+\frac{77.89}{2+n}+\right.\nonumber\\
&&\left.\frac{34.76}{3+n}-183.187 \left(\frac{1}{n}-s_1\right)+\frac{5120 s_1}{81 n}-65.43 \left(\frac{s_1}{n^2}+\frac{-\zeta (2)+s_2}{n}\right)\right)+\nonumber\\
&&\frac{32}{3} f^2 \left(-\frac{17}{72}+\frac{3-2 n-12 n^2+2 n^3+12 n^4}{27 n^3 (1+n)^3}+\frac{2 s_1}{27}+\frac{10 s_2}{27}-\frac{2
s_3}{9}\right)+\nonumber\\
&&502.4\left(-\frac{s_1}{n^3}+\frac{\zeta (2)-s_2}{n^2}-\frac{-\zeta (3)+s_3}{n}\right)\,. \nonumber
\end{eqnarray}
For the gluon-quark splitting functions we have
\begin{eqnarray}
&&\Delta p_{qg}^{(2)}=f \left(-\frac{1208}{n^5}+\frac{2313.84}{n^4}-\frac{1789.6}{n^3}+\frac{1461.2}{n^2}-\frac{2972.4}{n}+
\frac{439.8}{(1+n)^4}+\frac{2290.6}{(1+n)^3}+\frac{4672}{1+n}-\right.\nonumber\\
&&\frac{1221.6}{2+n}-\frac{18}{3+n}-\frac{278.32 s_1}{n}-\frac{90.26 \left(s_1{}^2+s_2\right)}{n}+825.4 \left(\frac{s_1}{n^2}+\frac{-\zeta (2)+s_2}{n}\right)+\nonumber\\
&&f \left(\frac{128}{3 n^5}-\frac{184.434}{n^4}+\frac{393.92}{n^3}-\frac{526.3}{n^2}+\frac{499.65}{n}-
\frac{61.116}{(1+n)^4}+\frac{358.2}{(1+n)^3}-\right.\nonumber\\
&&\frac{432.18}{1+n}-\frac{141.63}{2+n}-\frac{11.34}{3+n}+\frac{6.256 s_1}{n}+\frac{7.32 \left(s_1{}^2+s_2\right)}{n}-47.3 \left(\frac{s_1}{n^2}+\frac{-\zeta
(2)+s_2}{n}\right)+\nonumber\\
&&\left.\frac{0.7374 \left(-s_1{}^3-3 s_1 s_2-2 s_3\right)}{n}\right)-\frac{5.3 \left(-s_1{}^3-3 s_1 s_2-2 s_3\right)}{n}+\nonumber\\
&&\left.\frac{3.784 \left(s_1{}^4+6 s_1{}^2 s_2+3 s_2{}^2+8 s_1 s_3+6 s_4\right)}{n}\right)\,, \nonumber
\end{eqnarray}
\begin{eqnarray}
&&\Delta p_{gq}^{(2)}=\frac{92096}{27 n^5}-\frac{5328.018}{n^4}+\frac{4280}{n^3}-\frac{4046.6}{n^2}+\frac{6159}{n}-\frac{1050.6}{(1+n)^4}
-\frac{1701.4}{(1+n)^3}-\frac{3825.9}{1+n}+\nonumber\\
&&\frac{1942}{2+n}-\frac{742.1}{3+n}-\frac{1843.7 s_1}{n}+\frac{451.55 \left(s_1{}^2+s_2\right)}{n}-1424.8\left(\frac{s_1}{n^2}+\frac{-\zeta (2)+s_2}{n}\right)+\nonumber\\
&&f \left(-\frac{1024}{9 n^5}+\frac{236.3232}{n^4}-\frac{404.92}{n^3}+\frac{308.98}{n^2}-\frac{301.07}{n}+\frac{180.138}{(1+n)^4}
-\frac{253.06}{(1+n)^3}-\right.\nonumber\\
&&\frac{296}{1+n}+\frac{406.13}{2+n}-\frac{101.62}{3+n}+\frac{171.78 s_1}{n}-\frac{47.86 \left(s_1{}^2+s_2\right)}{n}-16.18\left(\frac{s_1}{n^2}+\frac{-\zeta
(2)+s_2}{n}\right)+\nonumber\\
&&\frac{16}{27} f \left(-\frac{12}{n}+\frac{10}{1+n}+\frac{2 }{1+n}\left(-\frac{1}{1+n}-s_1\right)-\frac{8 \text{ss}_1}{n}+\frac{6 \left(s_1{}^2+s_2\right)}{n}-\right.\nonumber\\
&&\left.\left.\frac{3 }{1+n}\left(\frac{1}{(1+n)^2}+\left(\frac{1}{1+n}+s_1\right){}^2+s_2\right)\right)-\frac{4.963 \left(-s_1{}^3-3 s_1 s_2-2
s_3\right)}{n}\right)+\nonumber\\
&&\frac{59.3 \left(-s_1{}^3-3 s_1 s_2-2 s_3\right)}{n}+\frac{5.143\left(s_1{}^4+6 s_1{}^2 s_2+3 s_2{}^2+8 s_1 s_3+6 s_4\right)}{n}\,.\nonumber
\end{eqnarray}

Finally, the polarized third-order gluon-gluon splitting function reads:
\begin{eqnarray}
&&\Delta
p_{gg}^{(2)}=4427.762+\frac{12096}{n^5}-\frac{22665}{n^4}+\frac{21804}{n^3}-\frac{23091}{n^2}+\frac{30988}{n}-
\frac{7002}{(1+n)^4}-\frac{1726}{(1+n)^3}-\nonumber\\
&&\frac{39925}{1+n}+\frac{13447}{2+n}-\frac{4576}{3+n}+2643.521 \left(\frac{1}{n}-s_1\right)-\frac{3801 s_1}{n}-\nonumber\\
&&13247 \left(-\frac{1}{1+n}\left(-\frac{1}{1+n}-s_1\right)-\frac{s_1}{n}\right)-12292 \left(\frac{s_1}{n^2}+\frac{-\zeta (2)+s_2}{n}\right)+\nonumber\\
&&f \left(-528.536-\frac{6128}{9 n^5}+\frac{2146.788}{n^4}-\frac{3754.4}{n^3}+\frac{3524}{n^2}-\frac{1173.5}{n}-\frac{786}{(1+n)^4}+\right.\nonumber\\
&&\frac{1226.2}{(1+n)^3}+\frac{2648.6}{1+n}-\frac{2160.8}{2+n}+\frac{1251.7}{3+n}-412.172 \left(\frac{1}{n}-s_1\right)+\frac{295.7 s_1}{n}-\nonumber\\
&&6746 \left(-\frac{1}{1+n}\left(-\frac{1}{1+n}-s_1\right)-\frac{s_1}{n}\right)-7932 \left(\frac{s_1}{n^2}+\frac{-\zeta (2)+s_2}{n}\right)+\nonumber\\
&&f \left(6.4607\, +\frac{7.0854}{n^4}-\frac{13.358}{n^3}+\frac{13.29}{n^2}-\frac{16.606}{n}+\frac{31.528}{(1+n)^3}+
\frac{32.905}{1+n}-\right.\nonumber\\
&&\left.\left.\frac{18.3}{2+n}+\frac{2.637}{3+n}-\frac{16}{9} \left(\frac{1}{n}-s_1\right)+\frac{0.21 s_1}{n}-16.944\left(\frac{s_1}{n^2}+\frac{-\zeta
(2)+s_2}{n}\right)\right)\right)\,. \nonumber
\end{eqnarray}
For completeness, we also include the polarized NNLO pure singlet contribution,
\begin{eqnarray}
&&\Delta p_{ps}^{(2)}=f \left(-\frac{344}{27} \left(\frac{24}{n^5}-\frac{24}{(1+n)^5}\right)-90.9198\left(-\frac{6}{n^4}+\frac{6}{(1+n)^4}\right)
-368.6\left(\frac{2}{n^3}-\frac{2}{(1+n)^3}\right)-\right.\nonumber\\
&&739 \left(-\frac{1}{n^2}+\frac{1}{(1+n)^2}\right)-1362.6\left(\frac{1}{n}-\frac{1}{1+n}\right)-81.5 \left(-\frac{6}{(1+n)^4}+\frac{6}{(2+n)^4}\right)+\nonumber\\
&&349.9 \left(\frac{2}{(1+n)^3}-\frac{2}{(2+n)^3}\right)+1617.4\left(\frac{1}{1+n}-\frac{1}{2+n}\right)-674.8 \left(\frac{1}{2+n}-\frac{1}{3+n}\right)+\nonumber\\
&&167.41 \left(\frac{1}{3+n}-\frac{1}{4+n}\right)-204.76 \left(-\frac{-\frac{1}{1+n}-s_1}{1+n}-\frac{s_1}{n}\right)+\nonumber\\
&&232.57 \left(\frac{s_1}{n^2}-\frac{\frac{1}{1+n}+s_1}{(1+n)^2}+\frac{-\zeta (2)+s_2}{n}-\frac{\frac{1}{(1+n)^2}-\zeta (2)+s_2}{1+n}\right)-\nonumber\\
&&12.61 \left(\frac{s_1{}^2+s_2}{n}-\frac{\frac{1}{(1+n)^2}+\left(\frac{1}{1+n}+s_1\right){}^2+s_2}{1+n}\right)+\nonumber\\
&&f \left(1.1741\left(-\frac{6}{n^4}+\frac{6}{(1+n)^4}\right)+13.287 \left(\frac{2}{n^3}-\frac{2}{(1+n)^3}\right)+45.482 \left(-\frac{1}{n^2}+\frac{1}{(1+n)^2}\right)+\right.\nonumber\\
&&49.13 \left(\frac{1}{n}-\frac{1}{1+n}\right)-0.8253\left(-\frac{6}{(1+n)^4}+\frac{6}{(2+n)^4}\right)+\nonumber\\
&&10.657 \left(\frac{2}{(1+n)^3}-\frac{2}{(2+n)^3}\right)-30.77 \left(\frac{1}{1+n}-\frac{1}{2+n}\right)-4.307 \left(\frac{1}{2+n}-\frac{1}{3+n}\right)-\nonumber\\
&&0.5094 \left(\frac{1}{3+n}-\frac{1}{4+n}\right)+9.517 \left(-\frac{1}{1+n}\left(-\frac{1}{1+n}-s_1\right)-\frac{s_1}{n}\right)+\nonumber\\
&&\left.1.7805\left(\frac{s_1{}^2+s_2}{n}-\frac{1}{1+n}\left(\frac{1}{(1+n)^2}+\left(\frac{1}{1+n}+s_1\right)
{}^2+s_2\right)\right)\right)-\nonumber\\
&&6.541\left(\frac{-s_1{}^3-3 s_1 s_2-2 s_3}{n}-\right.\nonumber\\
&&\left.\left.\frac{1}{1+n}\left(-\left(\frac{1}{1+n}+s_1\right){}^3-3 \left(\frac{1}{1+n}+s_1\right) \left(\frac{1}{(1+n)^2}+s_2\right)-2 \left(\frac{1}{(1+n)^3}+s_3\right)\right)\right)\right).\nonumber
\end{eqnarray}

\end{widetext}


%
%

\end{document}